\documentclass[aps,prb,twocolumn,superscriptaddress,nofootinbib,longbibliography,floatfix]{revtex4-2}

\usepackage{amsmath,amssymb,amsthm,bm,mathtools}
\usepackage{graphicx}
\usepackage{physics}
\usepackage{booktabs}
\usepackage{hyperref}
\usepackage{xcolor}
\usepackage{tikz}
\usepackage{placeins}

\usetikzlibrary{arrows.meta,positioning,calc,fit,backgrounds}

\hypersetup{
  colorlinks=true,
  linkcolor=blue!50!black,
  citecolor=blue!50!black,
  urlcolor=blue!50!black
}

\newcommand{\avg}[1]{\langle #1 \rangle}
\newcommand{\sign}{\ensuremath{\mathrm{sign}}}
\newcommand{\RSSE}{\ensuremath{\mathrm{RSSE}}}
\newcommand{\SSE}{\ensuremath{\mathrm{SSE}}}
\newcommand{\cv}{\ensuremath{\mathrm{cv}}}

\begin{document}



\title{Neural Autoregressive Control Variates for the Quantum Monte Carlo Sign Problem}

\author{Bei Qiao}
\affiliation{Beijing National Laboratory for Condensed Matter Physics and Institute of Physics, \\Chinese Academy of Sciences, Beijing 100190, China}
\affiliation{University of Chinese Academy of Sciences, Beijing 100049, China}

\author{Lei Wang}
\thanks{wanglei@iphy.ac.cn}
\affiliation{Beijing National Laboratory for Condensed Matter Physics and Institute of Physics, \\Chinese Academy of Sciences, Beijing 100190, China}

\date{\today}

\begin{abstract}
We train a pair of autoregressive models to construct zero-mean control variates to mitigate the sign problem in quantum Monte Carlo simulations. The two autoregressive networks are confined to the positive- and negative-sign sectors with strictly disjoint support, and each is exactly normalized over its sector. Their difference is therefore structurally zero-mean, providing an unbiased auxiliary observable whose correlation with the sign estimator controls the variance reduction. We implement the method within the stochastic series expansion framework, which we extend to frustrated lattices by developing an incremental loop-topology update. Sign-ergodic sampling is achieved through a twist channel, which is the unique sign-changing mechanism on non-bipartite lattices. We implement the control variates as autoregressive transformers with an end-of-sequence parity mask that enforces exact sign-sector resolution, while the incremental loop-count change and cumulative frustration parity are incorporated as topological features. On the triangular-lattice Heisenberg antiferromagnet, we benchmark the method in the small-$N$ limit. The control variate reduces the standard error of the average sign by up to an order of magnitude and that of the energy estimator by a factor of three to five, remaining effective even when the average sign drops below $10^{-3}$. This work lays out the framework and provides a proof-of-principle demonstration that autoregressive control variates can effectively mitigate the sign problem. Scaling to larger systems with physics-informed architectures is the subject of future work.
\end{abstract}

\maketitle
\tableofcontents

\section{Introduction}
The central difficulty in quantum many-body physics is the exponential growth of the Hilbert space with system size. Quantum Monte Carlo (QMC) is therefore one of the most important nonperturbative tools for correlated matter: it circumvents this exponential wall through stochastic sampling of the partition function, while retaining controllable statistical errors. For many physical systems of interest, the natural configuration weight is not positive---most notably fermionic systems in auxiliary-field representations and frustrated quantum magnets in worldline representations. The standard remedy is to sample configurations with weight $|W(x)|$ and treat the sign $s(x)\equiv \sign[W(x)]$ as part of the observable. Physical expectations then take the form
\begin{equation}
  \avg{O}
  =\frac{\avg{O\,s}_{|W|}}{\avg{s}_{|W|}},
  \label{eq:signreweight}
\end{equation}
where $\avg{\cdot}_{|W|}$ denotes the expectation under the absolute-weight measure $p(x)=|W(x)|/Z$, $Z=\sum_x|W(x)|$. The denominator $\avg{s}_{|W|}$ is the average sign. In the generic situation it decreases exponentially with space-time volume, causing an exponential deterioration of the signal-to-noise ratio. The sign problem is not restricted to a single model or algorithm. It is a fundamental obstacle across several areas of computational physics, including correlated fermion models in determinant or auxiliary-field QMC~\cite{Loh1990}, fermionic path-integral Monte Carlo for warm dense matter~\cite{Dornheim2019}, finite-density lattice QCD~\cite{Nagata2022}, nuclear auxiliary-field Monte Carlo~\cite{Alhassid1994}, and frustrated quantum magnets in worldline or stochastic-series-expansion representations~\cite{Henelius2000}. Although the microscopic origin of the negative weights differs from case to case, the numerical consequence is the same: physical quantities are obtained from cancellations between positive and negative contributions, and the cancellation becomes increasingly difficult to resolve statistically.

A major direction has therefore been to reformulate the problem so that the sign problem is absent or reduced by exploiting the basis dependence of the sign problem. In fermionic auxiliary-field QMC, antiunitary symmetries and related positivity conditions can guarantee nonnegative Monte Carlo weights, leading to broad classes of sign-problem-free interacting fermion models~\cite{WuZhang2005,Wei2016,LiJiangYao2016,LiYao2019}. In frustrated quantum magnets, one can sometimes work in cluster or dimer bases in which the effective weights are nonnegative, or at least have a much milder sign structure~\cite{Alet2016,Wessel2017,Wessel2018}. Beyond exploiting special symmetries or structures, the basis dependence can also be treated as an optimization problem: Shinaoka \emph{et al.}\ showed that the average sign in continuous-time QMC for cluster impurity problems can be substantially improved by choosing a nontrivial single-particle basis, for example one obtained by diagonalizing selected intracluster hoppings~\cite{Shinaoka2015}. Levy and Clark took the next step by treating single-particle basis rotations as variational degrees of freedom and searching for rotations that improve the average sign in Hubbard systems~\cite{LevyClark2021}. Since the average sign itself is generally not an efficient objective to optimize in severe sign-problem regimes, Hangleiter \emph{et al.}\ proposed a more general sign-easing framework based on efficiently computable measures of non-stoquasticity, optimized over local basis changes~\cite{Hangleiter2020}. These formulation-level approaches provide systematic ways to mitigate the sign problem by changing the Hamiltonian representation or sampling formulation.

Another line of work aims at a more broadly applicable mitigation strategy by embedding the fermionic problem into a family of auxiliary problems that are easier to simulate. A representative example is the fictitious-identical-particle construction of Xiong and Xiong, where the partition function is generalized by a real exchange-statistics parameter $\xi$ that interpolates between bosons, $\xi=1$, and fermions, $\xi=-1$~\cite{XiongXiong2022}. The practical idea is to simulate sign-free or sign-milder values of $\xi$, often on the $\xi\ge 0$ side, and extrapolate thermodynamic observables to the fermionic point. Related fictitious-particle or $\xi$-extrapolation ideas have been explored in path-integral simulations of warm dense matter, Hubbard-type lattice models, and normal liquid $^3$He~\cite{Dornheim2023FIP,Fan2026,MorresiGarberoglio2025}. Its strength, however, is also its conceptual vulnerability: the extrapolation assumes that the relevant thermodynamic quantities remain sufficiently analytic as functions of the auxiliary exchange-statistics parameter. Recent analyses based on Lee--Yang zeros indicate that this analyticity assumption can fail when zeros of the generalized partition function lie along or near the continuation path in the complex $\xi$ plane~\cite{He2025}. Thus, $\xi$-extrapolation methods may be powerful in favorable regimes, but they still rely on an auxiliary problem and an extrapolation assumption. This leaves open the need for strategies that work directly with the original sign-reweighted ensemble.

The present work follows a different route. We do not seek a sign-free basis or introduce an auxiliary statistics parameter for extrapolation. Instead, we keep the absolute-weight Monte Carlo sampling measure fixed and act only at the level of the estimator. Our goal is to reduce the variance of the sign-reweighted ratio estimator in Eq.~\eqref{eq:signreweight} through \emph{control variates} (CVs)---auxiliary functions whose expectation under the sampling measure is known exactly, so that subtracting them preserves the estimator's mean while reducing its variance.

Control variates have been developed for several related Monte Carlo problems. In lattice field theory with continuous field variables, this zero-mean property is often enforced through integration-by-parts or Schwinger-Dyson identities: Schwinger-Dyson relations provide analytically zero-mean control variates for scalar lattice field theory and lattice-fermion applications~\cite{BhattacharyaLawrenceYoo2024,Lawrence2024}. Neural versions of this idea parameterize the generating scalar or vector field by neural networks while preserving the zero-mean constraint by construction~\cite{BedaqueOh2023}. For discrete degrees of freedom, Lawrence and Yamauchi~\cite{LawrenceYamauchi2024} use a discrete-difference construction, with functions parameterized by extreme learning machines, to generate zero-mean control-variate bases and optimize their linear combination. In computer-graphics Monte Carlo integration, M\"uller \emph{et al.}~\cite{Muller2020} construct neural control variates whose integrals are tractable by construction, combined with a learned residual importance sampler. In all of these constructions, the zero-mean property is guaranteed either by an analytical identity of the theory or by an explicitly tractable integral of the learned approximator.

In this work we construct \emph{neural autoregressive control variates} for the QMC sign problem by training two models $q_+(x)$ and $q_-(x)$ confined to the even- and odd-parity sectors. The core requirement is \emph{exact normalization}: each model must be a properly normalized probability distribution, guaranteed by the autoregressive factorization. This ensures that the control variate $h(x) = [q_+(x) - q_-(x)] / |W(x)|$ has exactly zero expectation under the sampling measure $p(x)=|W(x)|/Z$---a mathematical guarantee that holds without estimation---thereby preserving the unbiasedness of the sign-reweighted estimator. The autoregressive factorization for discrete sequences plays the same role that normalizing flows play for continuous integrands in Ref.~\onlinecite{Muller2020}: both provide tractable exact normalization, but the autoregressive form is native to the variable-length operator strings produced by QMC. To maximize the variance reduction, we further impose \emph{strict sign-sector separation}: the supports of $q_+(x)$ and $q_-(x)$ are strictly disjoint, with each model confined to its respective parity sector. Because the two models never overlap, the learned control variate can capture as much of the sign structure as the model capacity allows, thereby maximizing the Pearson correlation $\rho$ with the sign estimator---the quantity that controls the variance-reduction factor $(1-\rho^2)$ [Eq.~(\ref{eq:varred})]. A practical consequence is that the training objective is entirely decoupled from the sign-problematic expectation: we use a standard cross-entropy loss that fits each model to the Markov-chain Monte Carlo (MCMC) samples within its sign sector. The training signal involves no sign estimator and remains stable regardless of how severe the sign problem is, while the variance reduction emerges only at evaluation time through the correlation between the learned control variate and the target observable.

This construction is designed for the Markov-chain setting of quantum Monte Carlo, where the sampling measure $p(x)$ is fixed by the Monte Carlo dynamics and cannot be replaced by a learned sampler. The two autoregressive models provide an auxiliary observable evaluated under this fixed measure, and the learning problem reduces to maximizing the correlation $\rho$ with the target sign estimator. Importantly, the control-variate framework itself is independent of the particular Monte Carlo backbone: it requires only a sign-ergodic sampler together with a representation in which the sign sectors can be tracked exactly.

\begin{figure*}[t]
\centering
\includegraphics[width=\textwidth]{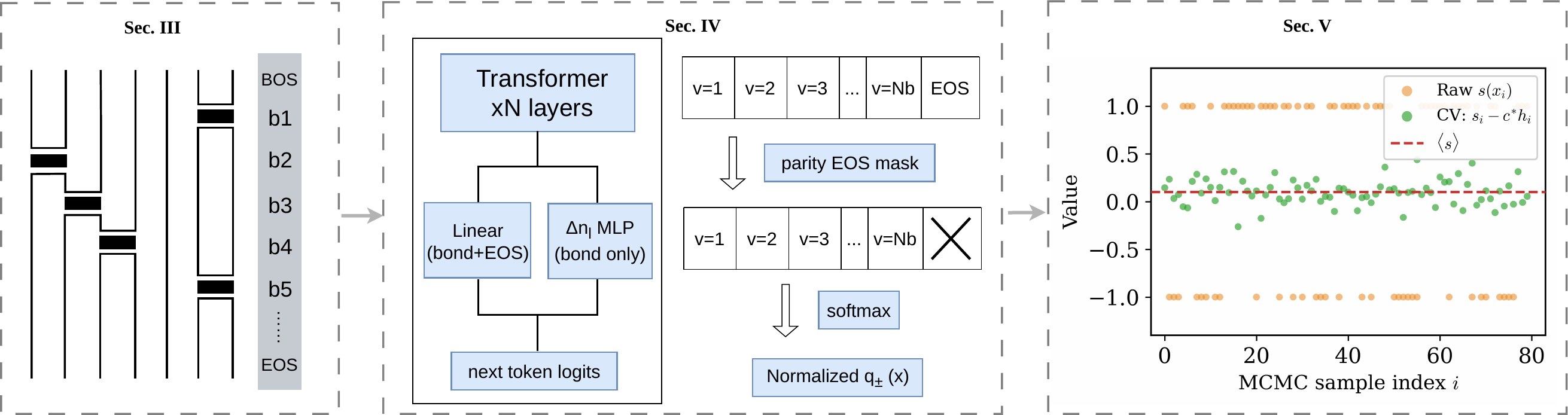}
\caption{\label{fig:pipeline}
Schematic overview of the workflow. Left (Sec.~\ref{sec:rsse}): an \SSE\ worldline configuration is converted into a compact uncolored operator string via the \RSSE\ resummation. Middle (Sec.~\ref{sec:network}): the operator string is processed by a causal transformer with a linear output head and a residual MLP that adds $\Delta n_l$-aware corrections to bond logits. A parity EOS mask enforces sign-sector resolution, and a softmax yields the normalized parity-resolved distributions $q_{\pm}(x)$. Right (Sec.~\ref{sec:results}): the resulting control variate $s_i - c^* h_i$ reduces the sign estimator's fluctuations around $\avg{s}$, while the raw sign $s(x_i)$ takes only values $\pm 1$.}
\end{figure*}

Figure~\ref{fig:pipeline} gives a schematic overview of the full workflow. The remainder of the paper is organized as follows. Section~\ref{sec:cv} introduces the general control-variate framework and the zero-mean construction for sign-sensitive estimators~\cite{Glynn2002,Shyamsundar2024}. Section~\ref{sec:rsse} describes the stochastic series expansion (SSE) backbone and the associated sign-ergodicity considerations. Section~\ref{sec:network} specifies the parity-aware autoregressive architecture and its physical constraints. Section~\ref{sec:results} presents benchmarks on frustrated triangular-lattice antiferromagnets. Section~\ref{sec:discussion} collects the numerical lessons and the remaining bottlenecks.

\section{Neural Control Variates for the Average Sign}
\label{sec:cv}

The general control-variate principle is as follows~\cite{Glynn2002,Assaraf1999}. Suppose one wishes to estimate $\avg{O}$ by Monte Carlo, and one has access to an auxiliary function $f(x)$ whose expectation $\avg{f}$ is known exactly. Then for any coefficient $c$,
\begin{equation}
  \avg{O - c\,(f-\avg{f})} = \avg{O},
\end{equation}
so subtracting $c\,(f-\avg{f})$ from the estimator preserves exactness. The optimal linear coefficient is
\begin{equation}
  c^* = \frac{\mathrm{Cov}(O,f)}{\mathrm{Var}(f)},
  \label{eq:cstar}
\end{equation}
which yields the reduced variance
\begin{equation}
  \mathrm{Var}(O - c^* f) = \mathrm{Var}(O)\,(1-\rho^2),
  \label{eq:varred}
\end{equation}
where $\rho$ is the correlation between $O$ and $f$. When $\avg{f}=0$ the improved estimator simplifies to $O - c\,f$.

This generalizes to multiple zero-mean control variates $\{f_i\}$, for which the optimal coefficients satisfy~\cite{Shyamsundar2024}
\begin{equation}
  \sum_{i} c_i^*\,\mathrm{Cov}(f_i,f_j) = \mathrm{Cov}(O,f_j),
  \label{eq:multicv}
\end{equation}
reducing to Eq.~(\ref{eq:cstar}) when there is a single control variate. In practice, the optimal CV coefficient $c^*$ should be estimated on data independent of the final Monte Carlo average to avoid finite-sample bias~\cite{Muller2020}.

\subsection{Sign-resolved models and zero-mean construction}
Let $|W(x)|$ denote the unnormalized configuration weight, $Z=\sum_x |W(x)|$ the partition sum, and $p(x)=|W(x)|/Z$ the normalized sampling measure from which the Markov chain draws configurations. Defining the sector weights $Z_{\pm}=\sum_{x:s(x)=\pm 1} |W(x)|$ with $Z=Z_++Z_-$, the average sign is
\begin{equation}
  \avg{s}=\frac{1}{Z}\sum_x |W(x)|\,s(x).
\end{equation}
When $\avg{s}\ll 1$, the variance of the raw estimator is dominated by cancellations between the two sectors.

We introduce two normalized models, $q_+(x)$ and $q_-(x)$, trained on the positive-sign and negative-sign samples, respectively. Exact support separation between the two sign sectors is enforced structurally by the parity-based end-of-sequence (EOS) mask discussed below. Their difference defines a zero-mean control variate:
\begin{equation}
  h(x)=\frac{q_+(x)-q_-(x)}{|W(x)|}.
  \label{eq:hx}
\end{equation}
The expectation vanishes because each $q_{\pm}$ is normalized over its sector, $\sum_x q_{\pm}(x)=1$, which is guaranteed exactly by the autoregressive factorization (Sec.~\ref{sec:network}):
\begin{align}
  \mathbb{E}_p[h]
  &=\frac{1}{Z}\sum_x |W(x)|\frac{q_+(x)-q_-(x)}{|W(x)|} \nonumber\\
  &=\frac{1}{Z}\sum_x q_+(x)-\frac{1}{Z}\sum_x q_-(x)=0.
  \label{eq:ehzero}
\end{align}

The improved estimator is therefore $\widehat{s}_{\cv}=N_{\mathrm{samp}}^{-1}\sum_{i}[s(x_i)-c^*\,h(x_i)]$, with the optimal coefficient $c^*=\mathrm{Cov}(s,h)/\mathrm{Var}(h)$ given by Eq.~(\ref{eq:cstar}). If the two sign-resolved models become exact, i.e.\ $q_{\pm}(x)=|W(x)|\mathbf{1}_{s(x)=\pm 1}/Z_{\pm}$, then $h(x)$ takes only two values, $1/Z_+$ or $-1/Z_-$, depending on the sign sector. Because $s(x)$ is likewise constant within each sector, the linear combination $s-c^*h$ becomes configuration independent:
\begin{equation}
  s(x)-c^*h(x)=\avg{s}\quad \text{for all } x,
  \label{eq:zerovar}
\end{equation}
so the denominator estimator reaches exact zero variance. The full derivation is given in Appendix~\ref{app:zerovar}. Two conditions are required for this limit: (i) exact normalization, $\sum_x q_{\pm}(x)=1$, which guarantees the zero-mean property, and (ii) exact parity resolution, $q_+(x)=0$ for all $x$ with $s(x)=-1$ and vice versa, so that each model has strictly disjoint support on the two sign sectors. Both conditions are enforced structurally by the autoregressive architecture described in Sec.~\ref{sec:network}.

It is instructive to compare this construction with the neural control variates of M\"uller \emph{et al.}~\cite{Muller2020} for continuous integration. That framework addresses non-negative integrands only: the control variate is decomposed as $g(x)=\bar{g}(x)\,G$, where a normalizing flow provides a normalized shape $\bar{g}$ and a separate neural network predicts the scalar integral value $G$, so that $\int g = G$ is known by construction. Because their residual correction is evaluated by an explicit importance-sampling estimator, the proposal density used to draw the residual samples can be learned, e.g.\ to approximate $|f-\alpha g|$. In the Markov-chain setting of the present work, the sampling measure is fixed and cannot be replaced by a learned sampler. This constraint also simplifies the problem: the zero-mean property follows from normalization of $q_{\pm}$ alone, without learning any scalar integral value. Moreover, the sign structure provides a discrete decomposition into two sectors, enabling the exact zero-variance limit of Eq.~(\ref{eq:zerovar})---a property that has no counterpart in the continuous-integration setting where the integrand varies continuously within both the positive and negative regions.

A practical subtlety arises from the scale of the control variate. Since each $q_{\pm}$ is normalized to unity while $|W(x)|$ is proportional to the partition function $Z$, the control variate $h(x)$ [Eq.~(\ref{eq:hx})] is of order $1/Z$ and becomes extremely small for large systems. The optimal coefficient $c^*\sim Z$ must compensate this scale, making a direct evaluation of Eq.~(\ref{eq:cstar}) or~(\ref{eq:multicv}) numerically unstable. We resolve this by extracting a log-space centering constant $C=-\mathrm{median}(\log q/|W|)$ from the training data and working with the rescaled variable $\tilde{h}(x)=e^C\,h(x)\sim O(1)$. The rescaled coefficient $\tilde{c}^*=\mathrm{Cov}(s,\tilde{h})/\mathrm{Var}(\tilde{h})$ is well conditioned, and the final CV correction $c^*h=\tilde{c}^*\tilde{h}$ is scale-invariant. This plays a role analogous to the learned scalar $G$ in Ref.~\onlinecite{Muller2020}, which absorbs the overall magnitude of the integrand.

\subsection{Numerator control variate and energy estimator}
When the observable $O(x)$ is strictly positive, the sign of the numerator integrand $O(x)\,s(x)$ is determined entirely by $s(x)$, and the same parity resolution used for the denominator applies. One trains a pair of sign-resolved models with $O$-reweighted loss, so that in the ideal limit
\begin{equation}
  q_{\pm}^{(O)}(x) \approx \frac{O(x)\,|W(x)|\,\mathbf{1}_{s(x)=\pm 1}}{Z_{\pm}^{(O)}},
\end{equation}
with $Z_{\pm}^{(O)}=\sum_{x:s=\pm 1} O(x)\,|W(x)|$. The corresponding control variate is
\begin{equation}
  h_O(x)=\frac{q_+^{(O)}(x)-q_-^{(O)}(x)}{|W(x)|}.
\end{equation}
As a concrete example, the energy per site in SSE takes the form $E/N=-\avg{n_h\,s}/(\beta N\avg{s})+\mathrm{const}$, where $n_h$ is the number of nonidentity operators in the operator string (see Sec.~\ref{sec:rsse}). Since $n_h>0$, the numerator $\avg{n_h\,s}$ falls into this category, and we apply the above construction with $O(x)=n_h(x)$.
Unlike the denominator, the numerator CV has no exact zero-variance limit: the residual variance is controlled by the asymmetry between $Z_+^{(O)}$ and $Z_-^{(O)}$ (see Appendix~\ref{app:zerovar} for the full analysis).

For observables that can take either sign, one can either extract $\sign[O(x)]$ and incorporate it into the parity classification---at the cost of modifying the EOS parity mask---or shift the observable by a constant $a$ so that $O'=O+a>0$. The shift adds $a$ to the ratio $R=\avg{O\,s}/\avg{s}$, since $\avg{O'\,s}/\avg{s}=R+a$, and is subtracted analytically from the final estimate. We do not pursue these extensions here, as the energy benchmark already exercises the essential features of the numerator CV.

Physical observables are typically obtained from ratios of the form $R=\avg{O\,s}/\avg{s}$. By the delta method~\cite{Cochran1977}, the variance of the ratio estimator to leading order is
\begin{align}
  \mathrm{Var}(R)&\approx\frac{1}{\avg{B}^2}\bigl[\mathrm{Var}(A)+R^2\,\mathrm{Var}(B)-2R\,\mathrm{Cov}(A,B)\bigr] \nonumber\\
  &=\frac{\mathrm{Var}(X)}{\avg{B}^2},
  \label{eq:deltavar}
\end{align}
where $A=O\,s$, $B=s$, and $X\equiv A-R\,B$. The covariance term is typically large and negative, because $A$ and $B$ share the sign factor $s$, and this natural cancellation is a significant part of the raw estimator's precision. Applying independent control variates to $A$ and $B$ reduces their individual variances but also reduces this covariance, so the improvement on the ratio can be much smaller than on either component alone. The correct strategy is to optimize the CV coefficients jointly via Eq.~(\ref{eq:multicv}) with $O=X$, where the numerator and denominator control variates form a $2\times 2$ system. The ratio $R$ is estimated from a training set independent of the test data. Since this initial estimate is itself noisy, one can iterate: apply the CV correction to obtain improved estimates of $A$ and $B$, re-estimate $R$, and recompute $c^*$, repeating until convergence. We adopt this joint optimization in the benchmarks of Sec.~\ref{sec:results}.

\section{RSSE Backbone and Sign-Ergodic Updates on Frustrated Lattices}
\label{sec:rsse}

The control-variate construction of Sec.~\ref{sec:cv} requires a Monte Carlo backbone that produces discrete operator-string configurations amenable to autoregressive modeling. The stochastic series expansion (\SSE)~\cite{Sandvik1999,Syljuasen2002} provides a natural framework: configurations are represented as ordered sequences of bond operators, which map directly onto token sequences for the neural network. Unlike worldline methods with a fixed imaginary-time discretization, the \SSE\ works directly with the Taylor expansion of $e^{-\beta H}$ and introduces no Trotter error.

We focus on the SU(2) antiferromagnetic Heisenberg model on frustrated lattices, where the sign problem is severe. It is useful to subdivide the Heisenberg interaction into its diagonal and off-diagonal parts in the standard basis of diagonal $z$ spin components. We define operators with two indices, $H_{a,b}$, with $a=1,2$ referring to diagonal and off-diagonal, respectively, and $b=1,\dots,N_b$ is the bond index. For each bond $b$ connecting sites $i$ and $j$,
\begin{align}
  H_{1,b} &= \frac{1}{4} - S_i^z S_j^z, \\
  H_{2,b} &= \frac{1}{2}(S_i^+ S_j^- + S_i^- S_j^+),
\end{align}
so that the full Hamiltonian is $H = -J\sum_{b=1}^{N_b}(H_{1,b}-H_{2,b}) + \text{const}$. The constant is irrelevant for the algorithm but will be included when calculating the energy. The reason for including the constant in $H_{1,b}$ is to make the series expansion positive-definite. Taylor-expanding the partition function gives
\begin{equation}
  Z=\mathrm{Tr}\,e^{-\beta H}
  =\sum_\alpha\sum_{n=0}^{\infty}(-1)^{n_{\mathrm{off}}}\frac{\beta^n}{n!}
   \sum_{S_n}
   \left\langle\alpha\left|\prod_{p=0}^{n-1} H_{a(p),b(p)}\right|\alpha\right\rangle,
  \label{eq:sse_taylor}
\end{equation}
where $S_n=[a(0),b(0)],\dots,[a(n-1),b(n-1)]$ is an operator string and $n_{\mathrm{off}}$ counts the off-diagonal operators. In practice, one truncates at cutoff $M$ and pads strings with identity operators $H_{0,0}=I$. For the Heisenberg model, every nonzero matrix element of $H_{a,b}$ equals $1/2$, so the sampling weight per configuration reduces to
\begin{equation}
  W(\alpha,S_M)
  =(-1)^{n_{\mathrm{off}}}\frac{(\beta/2)^n\,(M-n)!}{M!},
  \label{eq:sse_weight}
\end{equation}
where $n$ is the number of nonidentity operators. The cutoff $M$ is adjusted so that $M\gg\langle n\rangle$.

\subsection{RSSE resummation}

On frustrated lattices, however, the standard deterministic loop \SSE\ update is unable to cross between sign sectors~\cite{Henelius2000}. The sign of a colored configuration is
\begin{equation}
  s(C)=(-1)^{n_{\mathrm{off}}(C)},
  \label{eq:signconv}
\end{equation}
and the standard update consists of diagonal insertions/removals and loop flips. The diagonal update does not change $n_{\mathrm{off}}$. A loop flip toggles the spin assignment on a closed loop, reversing the operator type on every vertex the loop visits. A closed loop necessarily passes through an even number of operators, so $\Delta n_{\mathrm{off}}$ is even and the sign is unchanged. The colored \SSE\ dynamics is therefore locked within a single sign sector~\cite{Henelius2000}. Directed-loop updates based on the worm algorithm can in principle cross sign sectors by modifying the loop path itself rather than merely recoloring a frozen graph. However, on frustrated lattices such as the triangular antiferromagnet, the low-temperature physics is dominated by a highly degenerate manifold of minimally-frustrated plaquettes---each carrying two aligned and one anti-aligned spin---and the directed-loop dynamics tends to freeze within this manifold~\cite{Melko2007}, making it ineffective as a sign-ergodic sampler. Moreover, the asymmetric operator-string structure produced by the worm algorithm is unfavorable for autoregressive modeling.

To overcome this sign-sector obstruction while preserving the operator-string representation, we adopt the resummation-based \SSE\ (\RSSE), introduced by Desai and Pujari~\cite{Desai2021} for SU($N$) Heisenberg models on bipartite lattices, where the sign problem is absent and sign-sector ergodicity is not an issue. The key idea is to sum over all spin colorings compatible with a given uncolored loop graph, reducing the sampled object to a compact operator string whose weight depends only on the bond sequence and the topology of the loop graph. The original formulation was derived for bipartite lattices~\cite{Kaul2013}; for general SU($N$) the resummation does not extend to nonbipartite geometries, but the SU(2) case is exceptional: the bond operator is itself a singlet projector and does not require a sublattice-dependent spin reversal, so the resummation identity carries over to frustrated lattices. We exploit this to extend \RSSE\ to the SU(2) antiferromagnetic Heisenberg model on nonbipartite lattices, developing an incremental update scheme that tracks the uncolored loop topology and sign parity on the fly.

A colored configuration $(\alpha,S_M)$ is equivalent to an uncolored operator string together with a binary spin assignment on each loop: the bond sequence fixes the loop graph, and each of the $n_l$ loops can be independently colored in one of two ways, so $(\alpha,S_M)\leftrightarrow(\text{uncolored }S_M,\,\boldsymbol{\sigma}_{\mathrm{loop}})$ is a one-to-one correspondence. Summing over all $2^{n_l}$ colorings yields the uncolored weight
\begin{equation}
  |W(x)|
  =2^{n_l}
  \frac{(\beta/2)^{n_h}\,(M-n_h)!}{M!},
  \label{eq:wcloop}
\end{equation}
where $n_h$ is the number of nonidentity operators in the string and $n_l$ the number of loops in the uncolored vertex graph. The MCMC sampling is governed by the full weight, which depends on the cutoff $M$ through the combinatorial factor $(M-n_h)!/M!$. However, the $M-n_h$ identity operators in $S_M$ carry no physical information and merely serve as placeholders for the diagonal update. We therefore feed the neural network a dense representation $x=(b_1,b_2,\dots,b_{n_h})$ obtained by stripping all identity entries from $S_M$. In this compact form the cutoff $M$ drops out and the model learns only the physically meaningful bond sequence. A key advantage of working with uncolored operator strings is that they form a compact configuration space: any sequence of bond indices $(b_1,\dots,b_{n_h})$ corresponds to a valid physical configuration. By contrast, colored operator strings must satisfy spin-closure constraints---the initial and final spin states connected through the operator sequence must match---so the colored space contains many physically forbidden sequences. For autoregressive generative models this distinction is crucial: the uncolored representation allows every token sequence to carry nonzero probability, eliminating the need for constrained decoding or token masking~\cite{Geng2023,Willard2023} that would otherwise be required to enforce physicality.

\subsection{Incremental loop-topology update and twist channel}

\begin{figure}[t]
\centering
\includegraphics[width=\columnwidth]{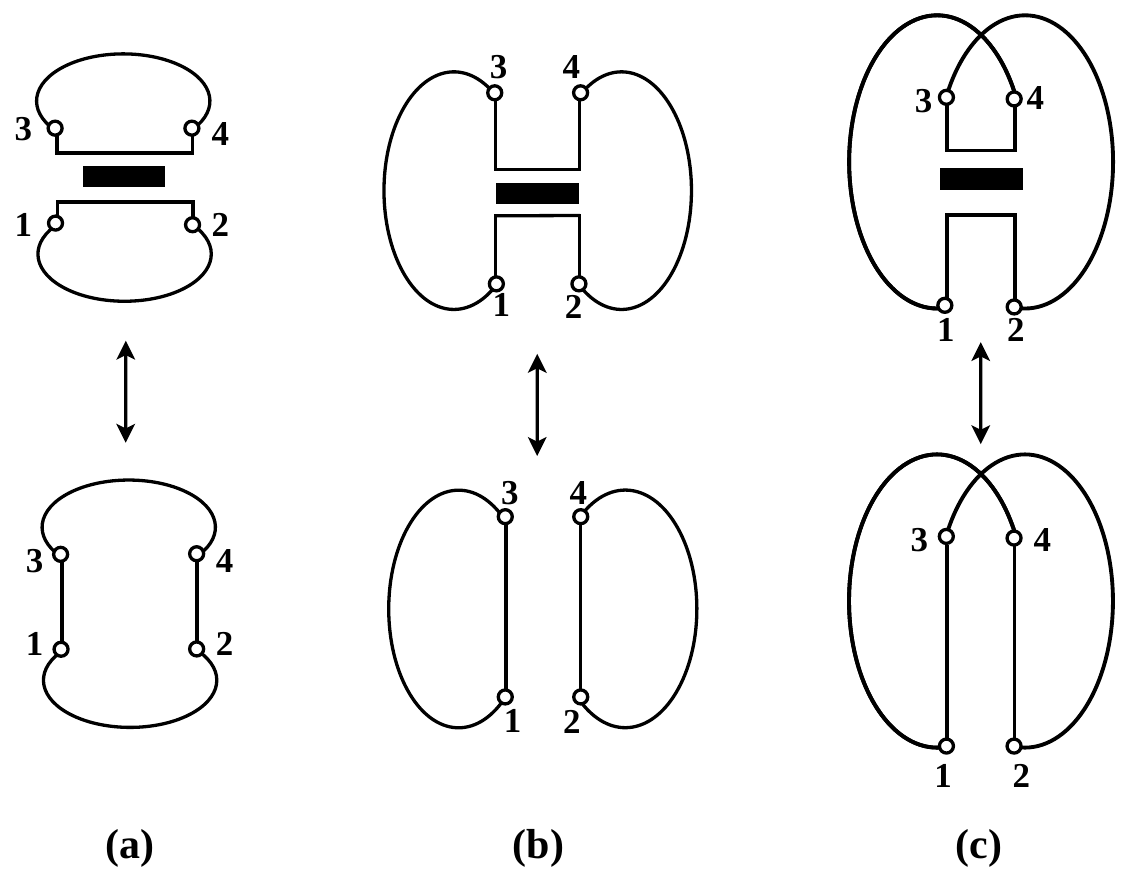}
\caption{\label{fig:topology}
Three topologically distinct reconnections when inserting an operator at a bond with legs 1,2 (bottom) and 3,4 (top). (a)~Split ($\Delta n_l\!=\!+1$): legs 1--2 and 3--4 connected within the same time slice; one loop splits into two. (b)~Merge ($\Delta n_l\!=\!-1$): legs 1--3 and 2--4 along worldlines; two loops merge into one. (c)~Twist ($\Delta n_l\!=\!0$): legs 1--4 and 2--3 cross-connected; loop count unchanged but internal routing rearranged. Only the twist channel changes the sign, and it is forbidden on bipartite lattices. Top: with operator (black bar). Bottom: after removal.}
\end{figure}

\RSSE\ overcomes the sign-sector obstruction of colored \SSE\ because its local move acts on the uncolored graph. Inserting or removing a single operator at a bond reconnects the four legs of the affected vertex. We label the legs as 1 and 2 at the bottom (sites $i$ and $j$ at imaginary time $\tau$), and 3 and 4 at the top (sites $i$ and $j$ at $\tau+1$), so that legs 1,3 share site $i$ and legs 2,4 share site $j$ via the worldline. There are exactly three topologically distinct reconnections (Fig.~\ref{fig:topology}):

\paragraph{Split $(\Delta n_l=+1)$.}
Legs 1--2 and 3--4 are connected within the same time slice, meaning all four legs belong to a single loop before insertion. Concretely, the intra-slice connections $1\!\leftrightarrow\!2$ and $3\!\leftrightarrow\!4$ combine with the worldline arcs $1\!\leftrightarrow\!3$ (site $i$) and $2\!\leftrightarrow\!4$ (site $j$) to form the single closed path $1$--$2$--$4$--$3$--$1$ [Fig.~\ref{fig:topology}(a), bottom]. The inserted operator cuts this loop into two. Since legs 1,3 share the worldline on site~$i$ they carry the same spin, and likewise legs 2,4 on site~$j$. The operator therefore sees identical entrance and exit spins on each site---it is a diagonal operator, $\Delta n_{\mathrm{off}}=0$, and the sign does not change.

\paragraph{Merge $(\Delta n_l=-1)$.}
Legs 1,3 on site $i$ form one loop and legs 2,4 on site $j$ form another [Fig.~\ref{fig:topology}(b), bottom]. The inserted operator bridges these two loops into one. By the same worldline continuity, $s_1=s_3$ and $s_2=s_4$, so the operator is again diagonal: $\Delta n_{\mathrm{off}}=0$, no sign change.

\paragraph{Twist $(\Delta n_l=0)$.}
Legs 1--4 and 2--3 are cross-connected, forming the closed path $1$--$4$--$2$--$3$--$1$. The loop count does not change, but the internal routing is rearranged. Consider the path from leg 1 to leg 4: it alternates between worldline segments and operators. Since leg 1 points downward in imaginary time and leg 4 also points downward, and each operator reverses the time direction, the path must traverse an even number of operators. This guarantees $s_1=s_4$ and similarly $s_2=s_3$. Identical spins on legs 1 and 4 (different sites) means the inserted operator must be off-diagonal. However, an off-diagonal operator requires opposite spins on legs 1 and 2, which is incompatible with the current spin configuration. The resolution is to flip all spins along one of the open paths (say, from leg 1 to leg 4). Because this path contains an even number of operators, each of which toggles between diagonal and off-diagonal under the flip, this global flip leaves the external $n_{\mathrm{off}}$ unchanged. The only net change comes from the inserted off-diagonal operator itself: $\Delta n_{\mathrm{off}}=1$, and the sign flips. 

The twist channel is therefore the unique mechanism by which \RSSE\ crosses between sign sectors. It is also specific to frustrated lattices, for a topological reason: the open path in a twist connects sites $i$ and $j$ across the bond via an alternating directed path that, as shown above, must pass through an even number of operators. On a bipartite lattice, $i$ and $j$ necessarily belong to different sublattices, and any path from one sublattice to the other traverses an odd number of bonds---a parity contradiction that forbids the twist channel entirely. This is the topological origin of the absence of the sign problem on bipartite lattices in the \SSE\ framework.

Since the twist channel is the sole sign-changing mechanism, the sign can be tracked incrementally on the uncolored graph: it flips if and only if the accepted move has $\Delta n_l=0$, without ever reconstructing a colored configuration.

The \RSSE\ Markov chain sweeps sequentially through all $M$ positions of the operator string. At each empty position, a random bond $b\in\{1,\dots,N_b\}$ is proposed for insertion, and the resulting change in loop count $\Delta n_l\in\{-1,0,+1\}$ is computed incrementally from the vertex graph. The move is accepted with probability
\begin{equation}
  P_{\mathrm{ins}}=\min\!\left(1,\;
    \frac{\beta N_b/2}{M-n_h}\;2^{\Delta n_l}\right),
  \label{eq:pinsert}
\end{equation}
where the first factor is the standard \SSE\ diagonal-update ratio~\cite{Sandvik1999} and $2^{\Delta n_l}$ accounts for the change in the number of loops under the resummation. At each occupied position, removal is proposed with acceptance probability
\begin{equation}
  P_{\mathrm{rem}}=\min\!\left(1,\;
    \frac{M-n_h+1}{\beta N_b/2}\;2^{\Delta n_l}\right),
  \label{eq:premove}
\end{equation}
where $\Delta n_l$ is the loop-count change upon removal. Detailed balance follows from $P_{\mathrm{ins}}/P_{\mathrm{rem}}=W(x')/W(x)\times(\text{proposal ratio})$ at each position. The three topological channels of Fig.~\ref{fig:topology} determine $\Delta n_l$: split gives $\Delta n_l=+1$, merge gives $\Delta n_l=-1$, and twist gives $\Delta n_l=0$.

To verify that the twist channel provides effective ergodic sampling across sign sectors, we measure the autocorrelation function of the sign itself. One MCMC sweep consists of sequentially visiting all $M$ positions in the operator string and proposing either an insertion (at empty positions) or a removal (at occupied positions) according to Eqs.~(\ref{eq:pinsert})--(\ref{eq:premove}). The sign autocorrelation function $C_s(\tau)=\langle s(t)\,s(t+\tau)\rangle - \langle s\rangle^2$ quantifies how quickly the Markov chain mixes between the two sign sectors. Appendix~\ref{app:acf} shows the measured autocorrelation functions for both the sign and the energy on the $3\times 3$ lattice at $\beta=3$, where $\avg{s}\approx 10^{-4}$ and the sign problem is extremely severe: the sign ACF decorrelates within a single sweep, rapidly reaching the $10^{-3}$ level, confirming that the twist channel provides rapid mixing between sign sectors even in this challenging regime.

It is important to emphasize that sign-ergodic sampling does not by itself reduce the variance of the sign estimator: the average sign $\avg{s}$ still decreases exponentially with system size and inverse temperature, and the signal-to-noise ratio deteriorates accordingly. Our incremental update scheme accelerates the \RSSE\ Markov chain on the triangular lattice by over two orders of magnitude compared to a naive implementation, but this speedup alone does not mitigate the sign problem---it merely makes the sign-ergodic sampling practical. What the \RSSE\ backbone provides is the prerequisite for the control-variate construction---a sampler that visits both sign sectors and produces compact uncolored operator strings on which the autoregressive model can act. Beyond this, the incremental loop-topology changes from the \RSSE\ update provide graph-topological features---in particular the running parity $\pi_t$ and the loop-count change $\Delta n_{l,t}$---that can be fed directly into the neural model to improve training (Sec.~\ref{sec:network}).

\section{Autoregressive Network Design}
\label{sec:network}

\subsection{Architecture and physical constraints}
The network acts on the compact uncolored operator string $x=(b_1,b_2,\dots,b_{n_h})$ rather than the padded \SSE\ string of length $M$. Each bond index $b_t\in\{1,\dots,N_b\}$ is mapped to a dense token ID $b_t+2$, so that the vocabulary consists of three special tokens (PAD, BOS, EOS) followed by $N_b$ operator tokens. The autoregressive sequence is $[\mathrm{BOS},b_1,\dots,b_{n_h},\mathrm{EOS}]$, padded to the batch maximum length with PAD tokens that are masked out from both the attention and the loss. Importantly, every such sequence corresponds to a valid physical configuration, ensuring that the autoregressive distribution assigns no probability weight to nonphysical states---a requirement for exact normalization over the physical configuration space.

For each target parity $\sigma\in\{0,1\}$ we model a normalized sequence distribution
\begin{equation}
  q_{\sigma}(x)=\prod_{t=1}^{n_h+1} q_{\sigma}(x_t\mid x_{<t}),
\end{equation}
where the final step predicts the EOS token. The network is a decoder-style causal transformer~\cite{Vaswani2017,Radford2019} with learned token and positional embeddings. Exact normalization over the physical configuration space is guaranteed by the autoregressive factorization, which requires a causal mask to prevent information leakage from future positions. Exact parity resolution---so that $q_+$ and $q_-$ have strictly disjoint support---is enforced by an EOS parity mask. A third mask on the sequence length ensures numerical stability. We describe each in turn.

\emph{Causal mask.}---The standard upper-triangular mask ensures that the hidden state at position $t$ attends only to positions $\le t$, defining a proper autoregressive factorization.

\emph{Parity EOS mask.}---When the running prefix parity $\pi_t$ is incompatible with the target sector $\sigma$, the EOS logit is set to $-\infty$, so the model can terminate only when $\pi_t=\sigma$. This guarantees that $q_+$ and $q_-$ have strictly disjoint support on the two sign sectors, which is required for the zero-variance limit of Eq.~(\ref{eq:zerovar}). The parity EOS mask is analogous to grammar-constrained decoding in large language models~\cite{Geng2023,Willard2023}, where the end-of-sequence token is masked at each autoregressive step unless the partially generated sequence satisfies a structural constraint. The difference here is that the constraint depends on a nonlocal property of the generated sequence---the cumulative frustration parity---rather than on local syntactic rules.

\emph{$n_h$ window mask.}---Rare configurations with very small or very large $n_h$ are exponentially underrepresented in the MCMC sample and poorly modeled, yet the control variate involves the ratio $q(x)/W(x)$, so even a modest error in $\log q$ for these configurations is exponentially amplified. To suppress this tail effect, we restrict the model to a window $n_{\min}\le n_h\le n_{\max}$: EOS is forbidden when the current sequence length is below $n_{\min}$, and EOS is forced when the length reaches $n_{\max}$. Configurations outside this window are assigned $q(x)=0$ and excluded from the CV evaluation. For the upper bound we set $n_{\max}=M$, the \SSE\ cutoff, which is always chosen larger than any $n_h$ that appears in the MCMC sample. This choice avoids an artificial probability spike at the boundary: because the autoregressive model generates tokens sequentially with no knowledge of how far $n_{\max}$ is, a tight upper cutoff would concentrate probability mass at the forced-termination point. Setting $n_{\max}=M$ ensures that no training or test sample actually reaches the boundary, so the model's learned distribution decays naturally before $n_{\max}$ and probability leakage beyond the physical support is minimized. A subtlety arises for boundary configurations whose parity at $n_{\max}$ may not match the target sector $\sigma$. To preserve the exact zero-mean property of the control variate, we evaluate such configurations under both $q_+$ and $q_-$ when computing $h(x)$, so that normalization is maintained exactly. This relaxes strict parity separation for a negligible fraction of extreme samples, but since $n_{\max}=M$ ensures that these configurations are effectively absent from the MCMC sample, the impact on variance reduction is negligible in practice.

The main architectural additions beyond a standard causal transformer are in the input embedding and the output head. The input at step $t$ is
\begin{equation}
  \tilde{e}_t=e_{b_t}+e_t^{\mathrm{pos}}+e_{\pi_t}^{\mathrm{par}}+e_{\Delta n_{l,t}}^{\mathrm{loop}},
  \label{eq:paremb}
\end{equation}
where $e_{\pi_t}^{\mathrm{par}}$ embeds the prefix parity $\pi_t\in\{0,1\}$ and $e_{\Delta n_{l,t}}^{\mathrm{loop}}$ embeds the incremental loop-count change $\Delta n_{l,t}\in\{-1,0,+1\}$ at position $t$. Both are deterministic functions of the operator prefix computed by the \RSSE\ incremental update and serve purely as auxiliary input features, so they do not alter the probability space. The parity embedding injects the cumulative frustration parity---a nonlocal graph-topological quantity---that the network would otherwise have to reconstruct from the bare bond sequence alone. The embedded sequence is then processed by a stack of $L$ causal transformer decoder layers, producing hidden states $h_t$ at each position.

On the output side, the base logits are computed by a linear projection $\ell_v = W_{\mathrm{out}}\,h_t$. Within a given $n_h$ sector the uncolored weight differs between configurations only through the factor $2^{n_l}$, so learning the correct next-token distribution requires the model to capture how the loop count changes for each candidate bond insertion. However, for a given prefix at position $t$, the model typically sees only the single next token that was actually sampled in the MCMC trajectory---it rarely encounters the same prefix paired with different candidate bonds, leaving insufficient training signal to learn the relative weights among different $\Delta n_l$ channels from data alone. To address this, we add a learned per-bond bias to the base logits, computed by a small multilayer perceptron (MLP) that receives the \emph{candidate} $\Delta n_l$ value and bond identity for each possible next bond $v$:
\begin{equation}
  \delta\ell_v = \mathrm{MLP}\bigl([h_t;\, e^{\mathrm{loop}}(\Delta n_{l,v}^{\mathrm{cand}});\, e^{\mathrm{bond}}(v)]\bigr),
\end{equation}
where $\Delta n_{l,v}^{\mathrm{cand}}\in\{-1,0,+1\}$ is the loop-count change that \emph{would} result from inserting bond $v$ at position $t+1$, computed by the same incremental vertex-graph construction used for the prefix parity, and $e^{\mathrm{bond}}(v)$ is a learned per-bond identity embedding that allows the MLP to distinguish bonds within the same $\Delta n_l$ class. The final logit for bond tokens is $\ell_v + \delta\ell_v$, while the EOS token retains only the base logit $\ell_{\mathrm{EOS}}$. The mean value of $\delta\ell_v$ across bond tokens effectively adjusts the relative weight between continuing the sequence and terminating at EOS. The MLP output layer is zero-initialized so that the bias correction starts at exactly zero and the model reduces to the baseline at initialization, ensuring stable training.

The conditional next-token probability is obtained by a softmax over the set of allowed tokens $\mathcal{V}_t(\sigma)$:
\begin{equation}
  q_\sigma(x_t=v\mid x_{<t})
  =\frac{\exp(\ell_v)}{\displaystyle\sum_{v'\in\mathcal{V}_t(\sigma)}\exp(\ell_{v'})},
  \label{eq:softmax}
\end{equation}
where $\mathcal{V}_t(\sigma)=\{$all operator tokens $\mid t < n_{\max}\}\cup\{\mathrm{EOS}\mid \pi_t=\sigma \text{ and } n_{\min}\le t\le n_{\max}\}$. Tokens outside $\mathcal{V}_t(\sigma)$ receive logit $-\infty$ and therefore zero probability. Because Eq.~(\ref{eq:softmax}) normalizes over $\mathcal{V}_t(\sigma)$ at every step, the autoregressive product $q_\sigma(x)=\prod_t q_\sigma(x_t\mid x_{<t})$ is exactly normalized over the physical configuration space restricted to the target parity sector.

\subsection{Training}
\label{subsec:training}

\paragraph{Loss functions.}
Both losses can be written as KL divergences, making the learning target explicit. For the denominator models, the target is $p_{\sigma}(x)\propto |W(x)|\,\mathbf{1}_{s(x)=\sigma}$, and the loss is
\begin{align}
  \mathcal{L}_{\mathrm{den}}
  &=\mathrm{KL}\bigl(p_{\sigma}\,\|\,q_{\sigma}\bigr)
  =\mathbb{E}_{x\sim p_{\sigma}}\!\left[\log\frac{p_{\sigma}(x)}{q_{\sigma}(x)}\right] \nonumber\\
  &=\mathrm{const}-\mathbb{E}_{x\sim p_{\sigma}}\log q_{\sigma}(x),
\end{align}
so that $q_\sigma$ learns the importance-sampling distribution $|W(x)|$ restricted to sign sector $\sigma$. The energy estimator requires the numerator $\avg{n_h\,s}$, since $E=-\avg{n_h\,s}/(\beta N\avg{s})+\mathrm{const}$. For the numerator models, the target is accordingly $p_{\sigma}^{(O)}(x)\propto n_h(x)\,|W(x)|\,\mathbf{1}_{s(x)=\sigma}$, and the loss is
\begin{align}
  \mathcal{L}_{\mathrm{num}}
  &=\mathrm{KL}\bigl(p_{\sigma}^{(O)}\,\|\,q_{\sigma}\bigr) \nonumber\\
  &=\mathrm{const}-\mathbb{E}_{x\sim p_{\sigma}^{(O)}}\log q_{\sigma}(x) \nonumber\\
  &=\mathrm{const}
  -\frac{\mathbb{E}_{x\sim p_{\sigma}}\bigl[n_h(x)\log q_{\sigma}(x)\bigr]}
  {\mathbb{E}_{x\sim p_{\sigma}}\bigl[n_h(x)\bigr]},
\end{align}
so that $q_\sigma$ learns the $n_h$-reweighted distribution, matching the numerator observable $\avg{n_h\,s}$. In practice, the denominator loss is the standard per-sequence negative log-likelihood $\mathcal{L}_{\mathrm{den}}=-N_{\mathrm{samp}}^{-1}\sum_{i}\log q_\sigma(x_i)$, while the numerator loss weights each sample by its operator count: $\mathcal{L}_{\mathrm{num}}=-N_{\mathrm{samp}}^{-1}\sum_{i} n_h(x_i)\log q_\sigma(x_i)$. The omitted normalization does not affect the minimizer.

\paragraph{Data augmentation.}
The compact operator string inherits a cyclic symmetry from the trace: all cyclic permutations of a string share the same weight $W$, loop count $n_l$, and sign. An autoregressive factorization breaks this symmetry by imposing a specific starting position. To mitigate this mismatch, each training sample is randomly cyclic-shifted at every epoch. In addition, for systems with spatial translation invariance (e.g.\ the $3\times 3$ lattice with periodic boundary conditions), we augment by applying random lattice translations, which permute the bond indices while preserving $W$, $n_l$, and the sign. For lattices where the spatial symmetry group is larger---specifically, the triangular torus with $L_x=L_y$ admits the dihedral group $D_6$ (six rotations and six reflections)---we further augment by applying random point-group operations, which act on the bond indices via the corresponding permutation representation. Because all three augmentations (cyclic shift, spatial translation, and point-group operation) change the prefix parity sequence, it must be recomputed for each augmented string. This is done on the fly by a Fortran module that incrementally builds the vertex graph and tracks the parity prefix, adding negligible overhead to the data pipeline.

\paragraph{Optimization.}
All models use $d_{\mathrm{model}}=128$, $L=4$ layers, $n_{\mathrm{head}}=4$, $d_{\mathrm{ff}}=512$, and dropout $0.1$. The residual MLP head uses a hidden dimension of $256$, a $\Delta n_l$ embedding dimension of $32$, and a bond embedding dimension of $32$. Models are optimized with Adam~\cite{Kingma2014} using gradient clipping at unit norm, batch size $128$, learning rate $5\times 10^{-5}$, and early stopping on validation loss with patience $20$ epochs. The training data consist of $2\times 10^5$ MCMC samples at the corresponding $\beta$, split into even- and odd-parity subsets, and each parity-resolved model is trained on its own subset of approximately $10^5$ samples. The sequence-length window parameters are set as follows: $n_{\min}$ is set to exclude the bottom ${\sim}0.01\%$ of rare short configurations in the training data, and $n_{\max}$ is set to the cutoff length $M$, which always exceeds the maximum observed $n_h$ in practice. Training samples with extreme $n_h$ values can be further clipped to a tighter range to reduce the influence of tail configurations on the learned distribution. The cross-entropy loss naturally assigns low probability to configurations outside the training range.

The computational overhead is modest compared to the variance reduction achieved. For each temperature point, training both autoregressive models and generating the $10^6$ effectively independent samples are comparable in computational cost. Efficient sampling is enabled by the incremental loop-topology update and the short autocorrelation times demonstrated in Fig.~\ref{fig:acf}. The forward pass of the trained models on these samples to construct the control variate incurs negligible additional cost. This overhead is far outweighed by the variance reduction: for the average sign, the up to 100-fold variance reduction (Table~\ref{tab:components}) is equivalent to a 100× reduction in the number of samples required to achieve the same statistical precision; for the energy estimator, the 3--5× standard error reduction (Table~\ref{tab:energy}) corresponds to a 9--25× reduction in required sample size. Even accounting for the per-temperature training cost, the control variate provides substantial net computational savings. This cost balance is expected to be even more favorable in methods or parameter regimes where producing decorrelated samples is the dominant computational bottleneck.
\section{Benchmarks}
\label{sec:results}

We benchmark the method on the triangular-lattice Heisenberg antiferromagnet at three system sizes: the 3-site triangular lattice, the $2\times 2$ lattice with periodic boundary conditions (4~sites, tetrahedron), and the $3\times 3$ lattice with periodic boundary conditions (9~sites). Exact diagonalization (ED) provides reference values for both $\avg{s}$ and the energy. Training, validation, and $c^*$ estimation are performed on independent \RSSE\ datasets at each $\beta$. This separation matters because using the test set to estimate $c^*$ would compromise the independence of the final evaluation. All test sets contain $10^6$ MCMC samples.

\subsection{3-site triangular lattice}

\begin{figure*}[t]
\centering
\includegraphics[width=\textwidth]{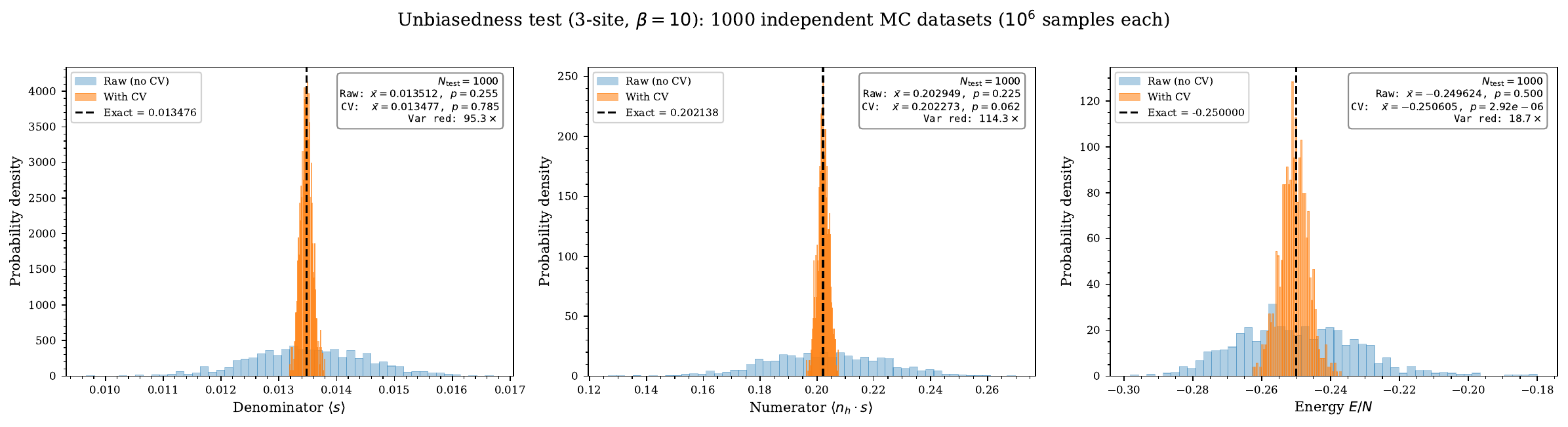}
\caption{\label{fig:bias_test}
Unbiasedness and finite-sample bias test at $\beta=10$ on the 3-site triangular lattice. Left and middle: distributions of the denominator $\avg{s}$ and numerator $\avg{n_h\,s}$ estimators over 1000 independent MCMC datasets of $10^6$ samples each, with and without control variates. Right: distribution of the energy ratio $E/N$ without jackknife bias correction. Dashed lines indicate exact values from ED. The CV estimators for the denominator and numerator are centered on the exact values, confirming unbiasedness at the component level, while achieving ensemble variance reductions of $95.3\times$ and $114.3\times$, respectively. The energy-ratio distribution shows a finite-sample ratio bias that becomes statistically visible after the CV suppresses the variance, despite achieving an ensemble variance reduction of $18.7\times$.}
\end{figure*}

\begin{figure}[t]
\centering
\includegraphics[width=\columnwidth]{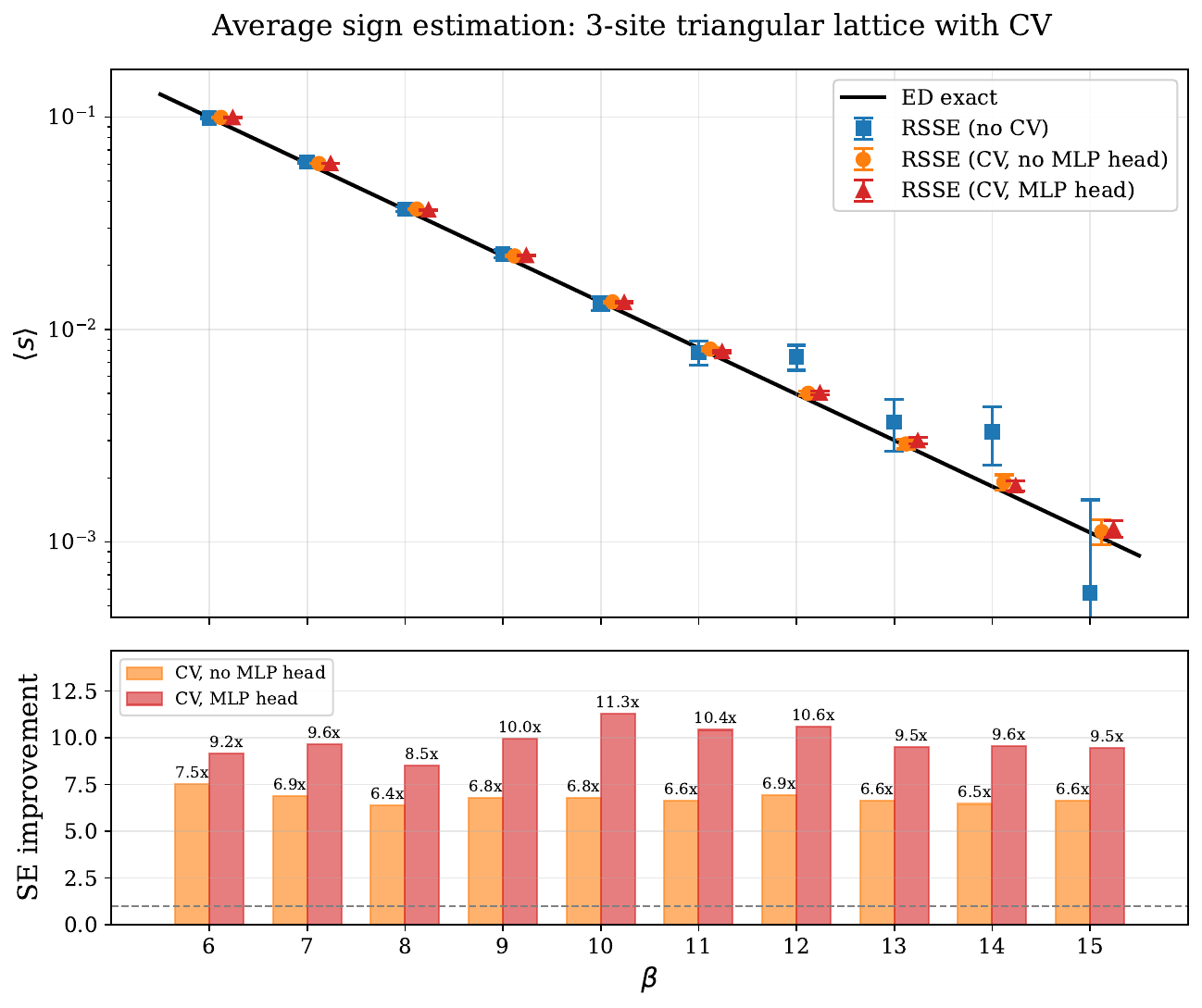}
\caption{\label{fig:sign_vs_beta}
Average sign $\avg{s}$ on the 3-site triangular lattice as a function of $\beta$, comparing the full model (with MLP head) and the ablated model (without MLP head). Top: expectation values of CV and raw estimators, compared with ED. Bottom: SE improvement factor. The MLP head provides consistent improvement across all $\beta$ values.}
\end{figure}

Table~\ref{tab:components} summarizes the component-level CV performance across $\beta=6$--$15$. The CV coefficients $c^*$ for the numerator and denominator are optimized independently at each temperature. Both models maintain high correlation ($\rho\approx 0.99$) and achieve variance reduction of $72$--$127\times$ for the denominator $\avg{s}$ and $61$--$119\times$ for the numerator $\avg{n_h s}$ across the full temperature range. The variance reduction remains strong even at high $\beta$ where the sign problem is severe and $\avg{s}$ drops below $10^{-2}$. Figure~\ref{fig:sign_vs_beta} shows the corresponding estimators compared with the ED reference. To verify that the CV introduces no systematic bias, Fig.~\ref{fig:bias_test} shows the distributions of both component estimators over 1000 independent MCMC datasets at $\beta=10$: the CV estimates for the denominator and numerator (left and middle panels) are centered on the exact ED values with dramatically reduced spread.

\begin{table*}[t]
\caption{\label{tab:components}
Component-level CV performance on the 3-site and $2\times 2$ (4-site) triangular lattices. Standard errors are shown in parentheses (last digits). VR: variance reduction. Test sets contain $10^6$ samples each.}
\begin{ruledtabular}
\begin{tabular}{cc@{\hspace{6pt}}cccc@{\hspace{6pt}}cccc}
 & & \multicolumn{4}{c}{Denominator $\avg{s}$} & \multicolumn{4}{c}{Numerator $\avg{n_h s}$} \\
\cline{3-6} \cline{7-10}
$\beta$ & $\avg{s}_{\mathrm{ED}}$ & Raw & CV & $\rho$ & VR & Raw & CV & $\rho$ & VR \\
\hline
\multicolumn{10}{c}{3-site triangular lattice} \\
\hline
6  & 0.09956 & 0.0992(10)  & 0.09947(11) & 0.994 & $84.2\times$ & 0.897(12)  & 0.8952(16) & 0.992 & $62.8\times$ \\
7  & 0.06039 & 0.0614(10)  & 0.06042(10) & 0.995 & $93.1\times$ & 0.645(14)  & 0.6315(16) & 0.994 & $83.7\times$ \\
8  & 0.03663 & 0.0369(10)  & 0.03653(12) & 0.993 & $72.5\times$ & 0.446(16)  & 0.4371(21) & 0.992 & $61.0\times$ \\
9  & 0.02222 & 0.0227(10)  & 0.02227(10) & 0.995 & $99.5\times$ & 0.304(18)  & 0.3023(20) & 0.994 & $88.1\times$ \\
10 & 0.01348 & 0.0133(10)  & 0.01341(9)  & 0.996 & $127\times$  & 0.192(20)  & 0.1997(23) & 0.994 & $81.0\times$ \\
11 & 0.00817 & 0.0078(10)  & 0.00787(10) & 0.995 & $109\times$  & 0.121(22)  & 0.1317(24) & 0.994 & $91.1\times$ \\
12 & 0.00496 & 0.0074(10)  & 0.00503(9)  & 0.996 & $113\times$  & 0.149(24)  & 0.0919(26) & 0.994 & $86.7\times$ \\
13 & 0.00301 & 0.0037(10)  & 0.00299(11) & 0.994 & $90.1\times$ & 0.075(26)  & 0.0572(24) & 0.996 & $119\times$  \\
14 & 0.00182 & 0.0033(10)  & 0.00184(10) & 0.995 & $91.3\times$ & 0.088(28)  & 0.0356(32) & 0.994 & $81.8\times$ \\
15 & 0.00111 & 0.0006(10)  & 0.00115(11) & 0.994 & $89.9\times$ & 0.017(30)  & 0.0276(29) & 0.995 & $110\times$  \\
\hline
\multicolumn{10}{c}{$2\times 2$ triangular lattice (4-site, tetrahedron)} \\
\hline
3   & 0.11003 & 0.1095(10)   & 0.11014(20)  & 0.979 & $23.8\times$ & 0.929(12)  & 0.9321(27) & 0.975 & $20.5\times$ \\
3.5 & 0.06452 & 0.0640(10)   & 0.06481(20)  & 0.979 & $23.7\times$ & 0.645(14)  & 0.6522(32) & 0.975 & $20.2\times$ \\
4   & 0.03821 & 0.0374(10)   & 0.03816(20)  & 0.980 & $25.4\times$ & 0.439(16)  & 0.4457(31) & 0.981 & $27.1\times$ \\
4.5 & 0.02281 & 0.0263(10)   & 0.02277(19)  & 0.982 & $28.4\times$ & 0.373(18)  & 0.3054(31) & 0.985 & $34.4\times$ \\
5   & 0.01370 & 0.0140(10)   & 0.01375(16)  & 0.988 & $40.5\times$ & 0.209(20)  & 0.2068(31) & 0.988 & $43.2\times$ \\
5.5 & 0.00826 & 0.0085(10)   & 0.00821(14)  & 0.989 & $47.8\times$ & 0.141(22)  & 0.1356(32) & 0.990 & $50.6\times$ \\
6   & 0.00499 & 0.0041(10)   & 0.00481(14)  & 0.990 & $48.5\times$ & 0.064(24)  & 0.0860(31) & 0.992 & $62.1\times$ \\
6.5 & 0.00302 & 0.0021(10)   & 0.00292(14)  & 0.990 & $50.1\times$ & 0.033(26)  & 0.0558(33) & 0.992 & $64.5\times$ \\
7   & 0.00183 & 0.0015(10)   & 0.00185(11)  & 0.993 & $77.1\times$ & 0.027(28)  & 0.0391(35) & 0.993 & $67.8\times$ \\
7.5 & 0.00111 & 0.00092(10) & 0.00100(12)  & 0.993 & $74.5\times$ & 0.021(30)  & 0.0167(38) & 0.992 & $62.9\times$ \\
8   & 0.00067 & 0.0007(10) & 0.00070(13)  & 0.991 & $58.0\times$ & 0.027(32)  & 0.0140(34) & 0.995 & $91.9\times$ \\
\end{tabular}
\end{ruledtabular}
\end{table*}

\begin{figure}[t]
\centering
\includegraphics[width=\columnwidth]{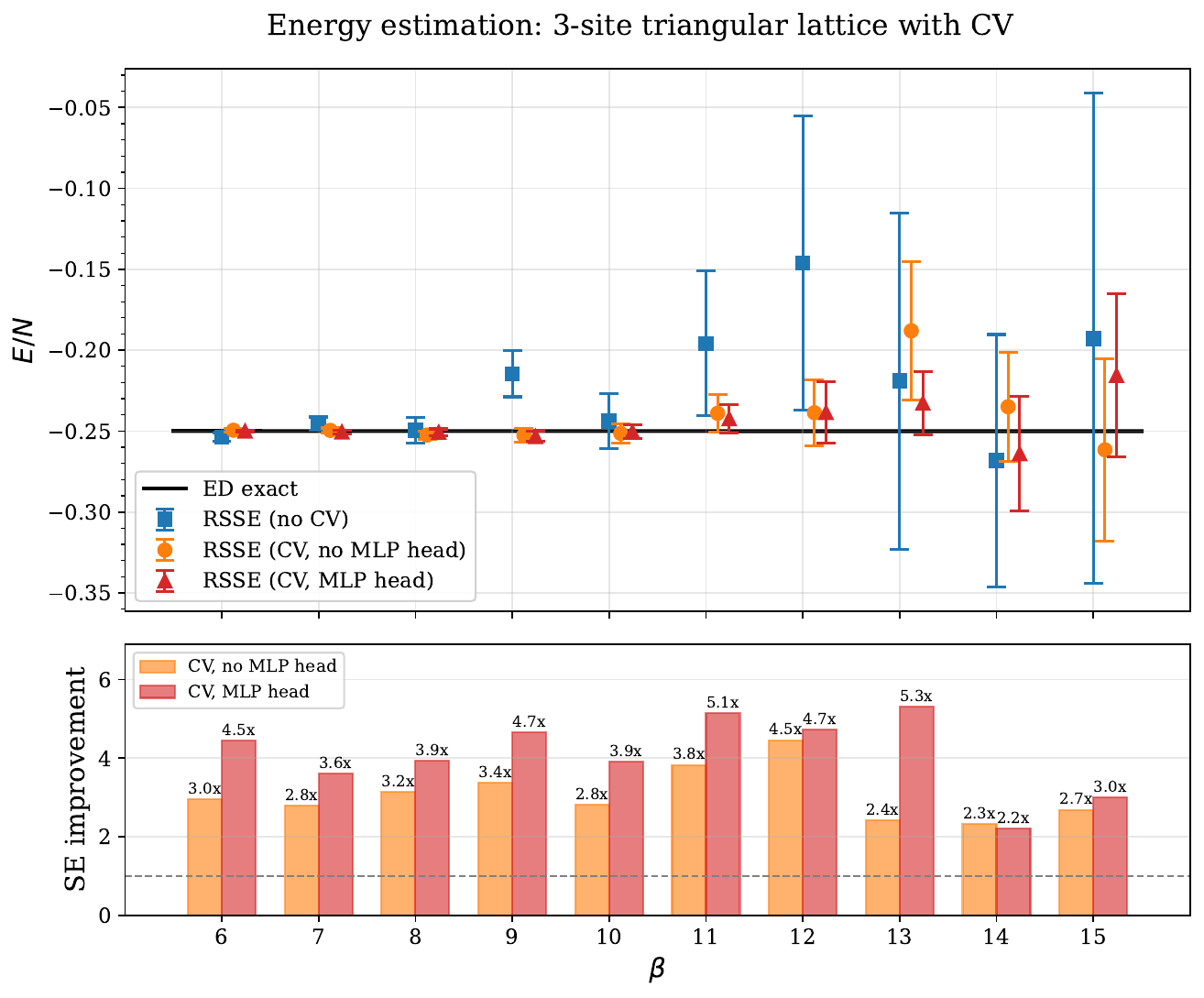}
\caption{\label{fig:energy_vs_beta}
Energy per site $E/N$ on the 3-site triangular lattice as a function of $\beta$, comparing the full model (with MLP head) and the ablated model (without MLP head). Top: CV estimator vs.\ raw estimator, compared with ED (black line). Error bars show one standard error. Bottom: SE improvement factor at each $\beta$. The MLP head yields a larger and more stable improvement, even at high $\beta$ where the sign problem is severe.}
\end{figure}

Physical observables require the ratio estimator. Because the energy is a nonlinear function of two correlated estimators, we use the jackknife method~\cite{WeigelJanke2010} to estimate its variance: the $N_{\mathrm{samp}}$ test samples are divided into $n_{\mathrm{bin}}$ bins, and the ratio is recomputed $n_{\mathrm{bin}}$ times with each bin removed in turn. The jackknife variance
\begin{equation}
  \mathrm{Var}_{\mathrm{JK}}(R)
  =\frac{n_{\mathrm{bin}}-1}{n_{\mathrm{bin}}}\sum_{k=1}^{n_{\mathrm{bin}}}
  \bigl(R_{(k)}-\bar{R}\bigr)^2
\end{equation}
naturally captures the correlation between numerator and denominator without assuming linearity.

As discussed in Sec.~\ref{sec:cv}, the improvement at the ratio level is generically smaller than the component-level gain: the numerator and denominator are naturally correlated through their shared sign factor, while independent CVs weaken this correlation. Table~\ref{tab:energy} and Fig.~\ref{fig:energy_vs_beta} show the energy per site as a function of $\beta$, comparing the CV and raw estimators against ED. The CV coefficients are optimized jointly via Eq.~(\ref{eq:multicv}). At low $\beta$ the sign problem is mild and both estimators agree with ED. As $\beta$ increases, the raw estimator develops large fluctuations and eventually becomes uninformative, while the CV estimator continues to track the ED curve with ${\sim}2.2$--$5.3\times$ SE improvement (Table~\ref{tab:energy}), weaker than the ${\sim}8$--$11\times$ component-level SE improvement.

\begin{table}[t]
\caption{\label{tab:energy}
Energy estimation on the 3-site and $2\times 2$ triangular lattices with joint $c^*$ optimization. Each row uses $10^6$ test samples. Central values are full-sample plug-in ratios (not jackknife-bias-corrected); standard errors are jackknife estimates. In the severe sign-problem regime, finite-sample bias may be comparable to statistical error.}
\begin{ruledtabular}
\begin{tabular}{ccccc}
$\beta$ & $E/N_{\mathrm{ED}}$ & $E/N$ (raw) & $E/N$ (CV) & SE impr. \\
\hline
\multicolumn{5}{c}{3-site triangular lattice} \\
\hline
6  & $-0.2499$ & $-0.254(2)$   & $-0.2499(5)$  & $4.45\times$ \\
7  & $-0.2500$ & $-0.245(4)$   & $-0.251(1)$   & $3.62\times$ \\
8  & $-0.2500$ & $-0.250(7)$   & $-0.251(2)$   & $3.93\times$ \\
9  & $-0.2500$ & $-0.21(1)$    & $-0.253(3)$   & $4.65\times$ \\
10 & $-0.2500$ & $-0.24(1)$    & $-0.250(4)$   & $3.91\times$ \\
11 & $-0.2500$ & $-0.20(4)$    & $-0.242(8)$   & $5.15\times$ \\
12 & $-0.2500$ & $-0.15(9)$    & $-0.24(1)$    & $4.73\times$ \\
13 & $-0.2500$ & $-0.22(10)$   & $-0.23(1)$    & $5.31\times$ \\
14 & $-0.2500$ & $-0.27(7)$    & $-0.26(3)$    & $2.21\times$ \\
15 & $-0.2500$ & $-0.19(15)$   & $-0.22(5)$    & $3.00\times$ \\
\hline
\multicolumn{5}{c}{$2\times 2$ triangular lattice (4-site, tetrahedron)} \\
\hline
3   & $-0.3291$ & $-0.328(4)$   & $-0.331(1)$  & $3.61\times$ \\
3.5 & $-0.3450$ & $-0.344(6)$   & $-0.345(1)$  & $3.98\times$ \\
4   & $-0.3560$ & $-0.36(1)$    & $-0.357(2)$  & $3.71\times$ \\
4.5 & $-0.3631$ & $-0.34(1)$    & $-0.367(4)$  & $3.53\times$ \\
5   & $-0.3676$ & $-0.37(2)$    & $-0.375(6)$  & $4.23\times$ \\
5.5 & $-0.3705$ & $-0.37(3)$    & $-0.36(1)$   & $3.68\times$ \\
6   & $-0.3722$ & $-0.27(9)$    & $-0.36(2)$   & $4.04\times$ \\
6.5 & $-0.3733$ & $-0.24(21)$   & $-0.38(5)$   & $4.25\times$ \\
7   & $-0.3740$ & $-0.29(33)$   & $-0.42(6)$   & $4.99\times$ \\
7.5 & $-0.3744$ & $-0.44(28)$   & $-0.45(11)$  & $2.44\times$ \\
8   & $-0.3746$ & $-0.11(87)$   & $-0.30(21)$  & $4.03\times$ \\
\end{tabular}
\end{ruledtabular}
\end{table}

An additional subtlety is the finite-sample bias of the ratio estimator. Since $E/N$ is obtained as a ratio of two sample means, the plug-in estimator $\hat{A}/\hat{B}$ has an $O(1/N)$ bias in general, while its statistical fluctuation scales as $O(N^{-1/2})$. Thus the estimator is consistent and asymptotically unbiased for $\avg{B}\neq 0$. We use the jackknife not as a bias-removal procedure, but to estimate the standard error of the nonlinear ratio estimator. Figure~\ref{fig:bias_test} (right panel) illustrates this effect at $\beta=10$: over 1000 independent datasets, the CV energy ratio has mean $\bar{x}=-0.250605$ compared with the exact value $-0.250000$ and achieves an $18.7\times$ ensemble variance reduction, corresponding to a ${\sim}4.3\times$ standard-error improvement. The small offset is statistically significant ($p=2.92\times 10^{-6}$), but it is a generic finite-sample property of ratio estimators and is not introduced by the control variate. The CV merely makes it visible by suppressing the variance. The standard jackknife bias correction subtracts the leading-order bias estimate,
\begin{equation}
  R_{\mathrm{JK}} = n_{\mathrm{bin}}\,\bar{R}
    - (n_{\mathrm{bin}}-1)\,\overline{R_{(k)}},
\end{equation}
where $\bar{R}$ is the full-sample ratio and $\overline{R_{(k)}}$ is the mean of the leave-one-bin-out ratios. This correction removes the $O(1/N_{\mathrm{samp}})$ bias when the denominator $\avg{s}$ is well separated from zero. In the severe sign-problem regime, however, $\avg{s}\to 0$ and the ratio $R=\avg{n_h\,s}/\avg{s}$ develops higher-order singularities that the leading-order jackknife cannot capture. Residual finite-sample bias can therefore persist even after correction, and the narrowed CV distribution can make such small offsets more visible. Consequently, all expectation values reported in this work are computed as the uncorrected full-sample ratio $\bar{R}$, without applying the leading-order bias estimate $R_{\mathrm{JK}}$.

\subsection{\texorpdfstring{$2\times 2$}{2x2} triangular lattice (tetrahedron)}

\begin{figure}[t]
\centering
\includegraphics[width=\columnwidth]{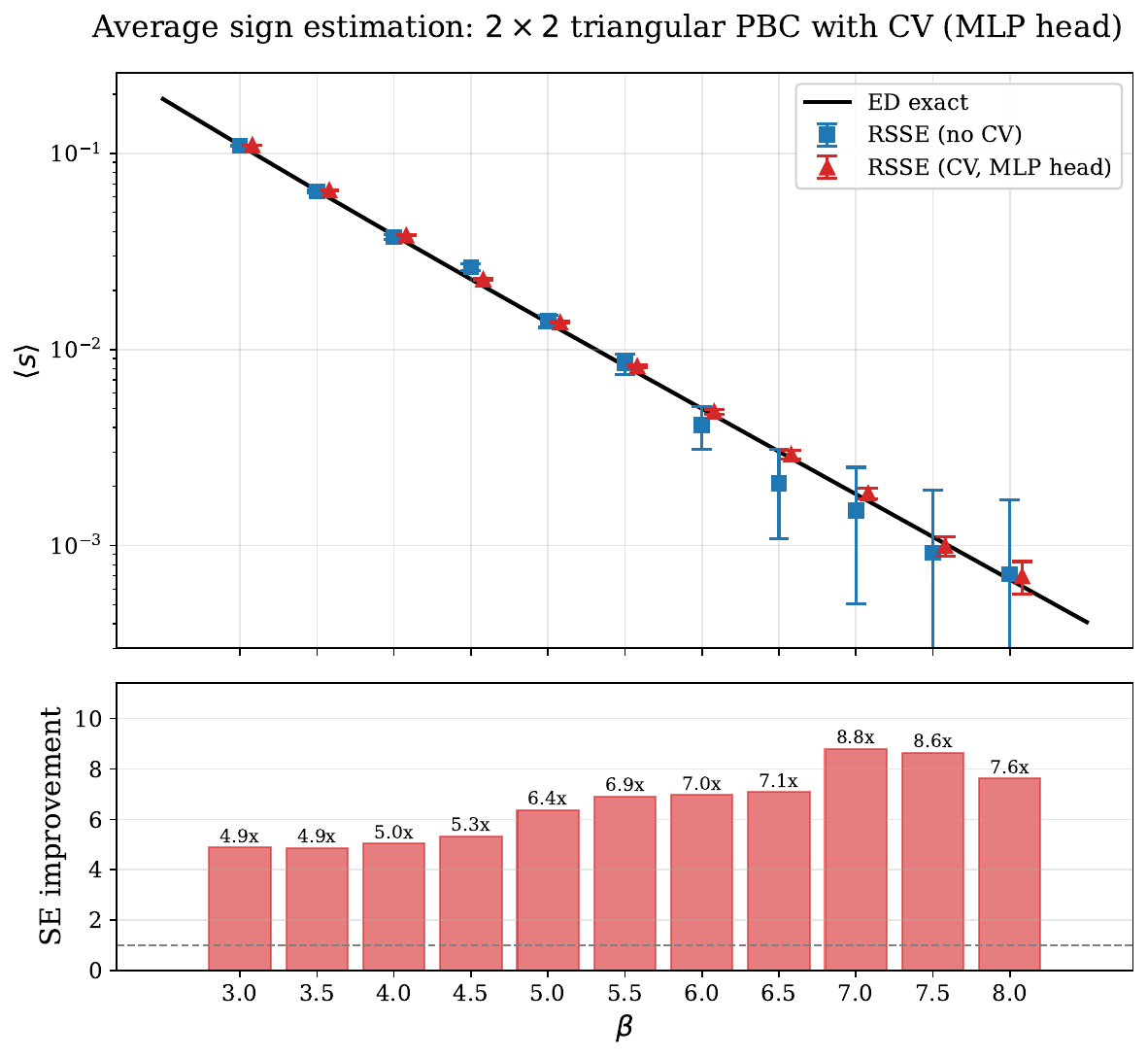}
\caption{\label{fig:2x2_sign}
Average sign $\avg{s}$ on the $2\times 2$ triangular lattice (tetrahedron) as a function of $\beta$. Top: expectation values of CV and raw estimators, compared with ED. Bottom: SE improvement factor. The CV achieves ${\sim}5$--$9\times$ SE improvement across $\beta=3$--$8$.}
\end{figure}

\begin{figure}[t]
\centering
\includegraphics[width=\columnwidth]{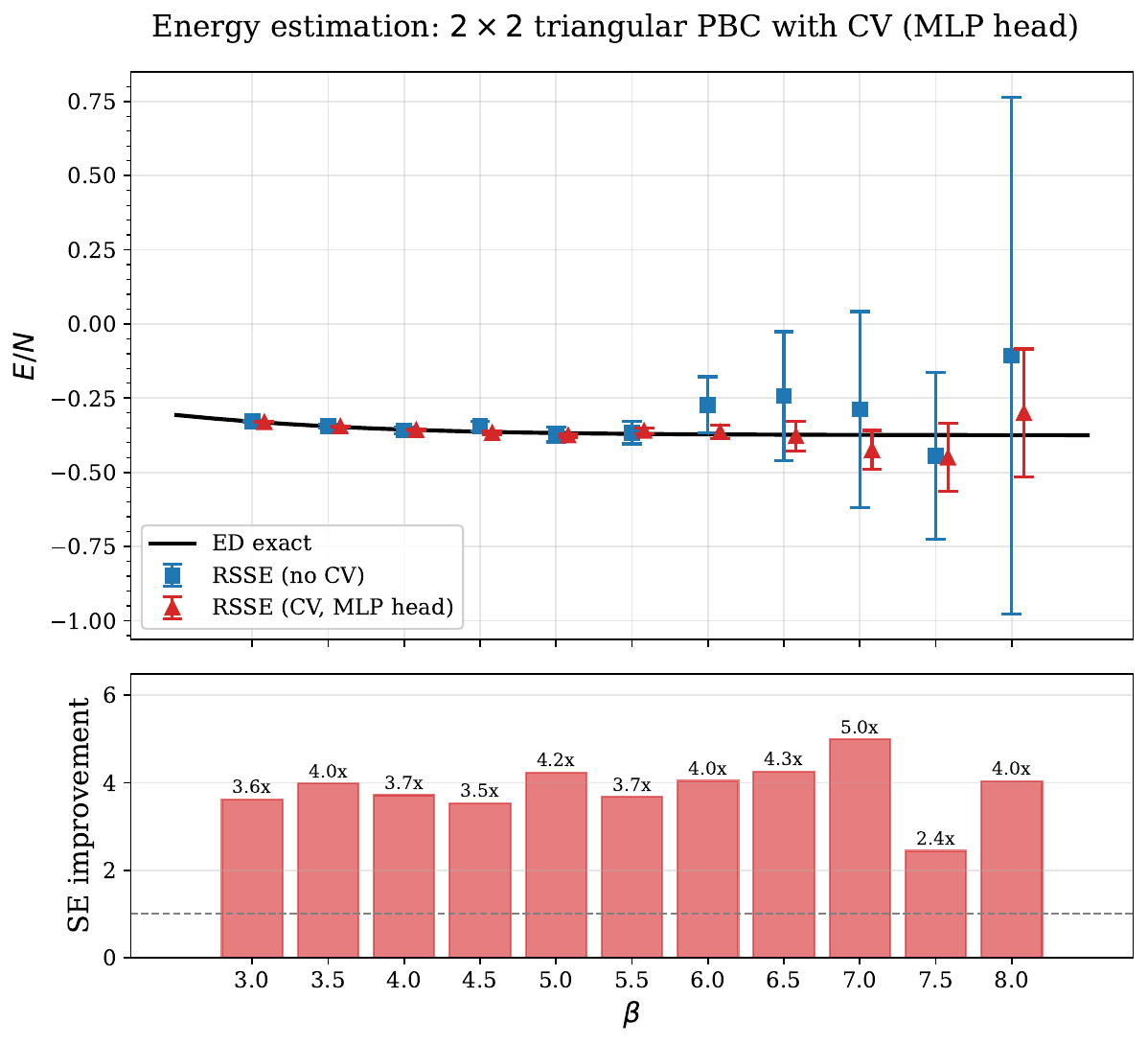}
\caption{\label{fig:2x2_energy}
Energy per site on the $2\times 2$ triangular lattice as a function of $\beta$. Top: CV (full model with MLP head) vs.\ raw estimator, compared with ED. Bottom: SE improvement factor, achieving ${\sim}2.4$--$5\times$ improvement.}
\end{figure}

To test the method on a different geometry, we repeat the benchmarks on the $2\times 2$ triangular lattice with periodic boundary conditions (4~sites, equivalent to a tetrahedron). Table~\ref{tab:components} shows component-level performance across $\beta=3$--$8$: the CV achieves variance reduction of $23.8$--$77.1\times$ for the denominator and $20.2$--$91.9\times$ for the numerator, with correlation $\rho\approx 0.98$--$0.99$. The variance reduction is somewhat lower than on the 3-site lattice, reflecting the increased difficulty of modeling the sign structure on a more frustrated geometry. Figure~\ref{fig:2x2_sign} shows the average-sign estimation: the CV maintains effective variance reduction even as $\avg{s}$ drops below $10^{-3}$. Table~\ref{tab:energy} and Figure~\ref{fig:2x2_energy} show the corresponding energy estimation, with ${\sim}2.4$--$5.0\times$ SE improvement on $E/N$.

\begin{figure*}[t]
\centering
\includegraphics[width=0.95\textwidth]{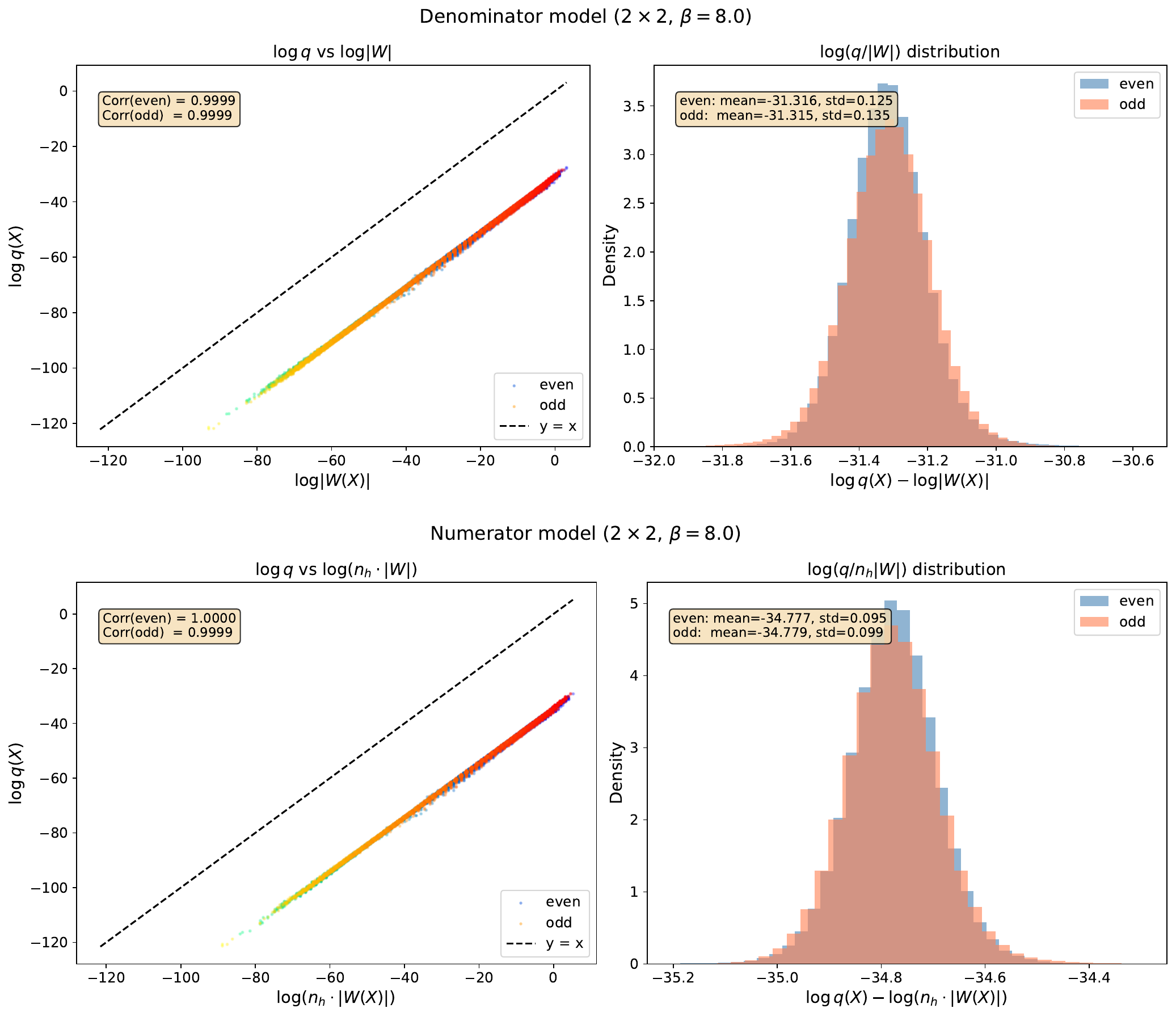}
\caption{\label{fig:logqlogw}
Parity-resolved comparison between the learned log-probability and the target log-weight on the $2\times 2$ triangular lattice at $\beta=8$. Top: denominator model, $\log q(x)$ vs.\ $\log |W(x)|$. Bottom: numerator model, $\log q(x)$ vs.\ $\log(n_h\,|W(x)|)$. Left panels show the scatter plot for even and odd parity sectors; right panels show the distribution of $\log(q/|W|)$ (denominator) and $\log(q/n_h |W|)$ (numerator). The strong linear relation confirms that the coarse dependence on $(n_h,n_l)$ is easily learned, but the nontrivial part for variance reduction is the residual sign-resolved structure within each sector, reflected in the narrow but finite width of the $\log(q/|W|)$ distributions. The vertical offset from the diagonal $y=x$ in the scatter plots equals $-\log Z_\pm$ for the denominator and $-\log Z_\pm^O$ for the numerator, reflecting that $|W(x)|$ and $n_h|W(x)|$ are unnormalized while $q(x)$ is a normalized probability distribution within each parity sector.
}
\end{figure*}

\subsection{\texorpdfstring{$3\times 3$}{3x3} triangular lattice}

To probe the scaling to larger systems, we apply the method to the $3\times 3$ triangular lattice with periodic boundary conditions (9~sites, 27~bonds). The autoregressive model has the same architecture as for the smaller lattices but now operates on a larger vocabulary ($N_b=27$ vs.\ $N_b=3$ or $6$ for the smaller lattices). The sequence length range remains comparable across all benchmarks ($n_{h,\text{max}}\approx 60$), so the primary challenge is the expanded vocabulary. On this larger system, the variance reduction is substantially weaker than on the 3- and 4-site lattices: the correlation between the control variate and the sign estimator drops to $\rho\approx 0.9$ (compared to $\rho\approx 0.99$ on smaller systems), and the standard error improvement on the energy estimator is modest.

The degradation in performance as system size increases is expected and does not reflect a fundamental limitation of the autoregressive CV framework. Importantly, the model capacity (embedding dimension $d_{\text{model}}=128$, number of layers $L=4$) is held fixed across all system sizes in this proof-of-principle study. As the system size grows, the configuration space expands exponentially while the model capacity remains constant, leading to underfitting: the fixed-capacity model cannot capture the increasingly complex sign structure. This is a standard requirement in neural-network applications---model capacity must scale with problem complexity. We discuss the implications for scaling in Sec.~\ref{sec:discussion}.

\subsection{MLP head ablation}

To quantify the contribution of the residual MLP head, we compare the full model against an ablated variant in which the MLP head is removed and the transformer output is projected directly to logits.

The results are shown in Figures~\ref{fig:sign_vs_beta} and~\ref{fig:energy_vs_beta} for the 3-site lattice. For the average-sign estimator, the MLP head improves the SE reduction from ${\sim}6.6$--$7.5\times$ (ablated) to ${\sim}8.5$--$11.3\times$ (full model) across $\beta=6$--$15$, a relative gain of ${\sim}30$--$50\%$. For the energy, the improvement is from ${\sim}2.3$--$4.5\times$ to ${\sim}2.2$--$5.3\times$.

The consistent improvement across all $\beta$ values confirms that the $\Delta n_l$ information is a key feature for resolving the sign structure, and that injecting it explicitly is more effective than relying on the transformer to learn it from data alone. The ablation results show that the transformer alone struggles to recover this graph-level information, and that even the simple explicit injection of $\Delta n_l$ yields a substantial gain. This suggests that enhancing the model's ability to capture graph structure---e.g.\ through graph neural networks or message-passing layers that encode lattice connectivity directly---is a promising direction for further improving the variance reduction.

We note that the ablation study focuses on the MLP head, which provides the largest single architectural contribution (${\sim}30$--$50\%$ relative gain). A complete ablation of all components---including the prefix parity embedding and cyclic and spatial augmentation---was not performed in this proof-of-concept study. The primary objective here is to demonstrate the feasibility of the autoregressive CV framework and identify the key architectural elements that enable effective variance reduction, rather than to claim an optimal or production-ready architecture. Future work on scaling to larger systems will require systematic ablation and hyperparameter tuning to determine the most efficient combination of inductive biases for a given target problem.

\subsection{Learning target analysis}

The \RSSE\ weight depends only on two coarse variables,
\begin{equation}
  \log |W(x)|=n_l(x)\log 2+n_h(x)\log(\beta/2)-\log(n_h!),
\end{equation}
so the target distribution itself has a simple structure. However, the autoregressive model does not learn $W(x)$ directly. It learns the conditional next-token probabilities $q(x_t\mid x_{<t})$, whose product must reproduce the joint distribution. The joint log-probability $\log q(x)=\sum_{t}\log q(x_t\mid x_{<t})$ accumulates errors from each step, so the effective modeling difficulty grows with the sequence length $n_h$. This is why the within-sector conditional
\begin{equation}
  \frac{q_{\pm}(x)}{W(x)}
  =\frac{q_{\pm}(n_h,n_l)}{W(n_h,n_l)}\, q_{\pm}(x\mid n_h,n_l)
\end{equation}
is the CV-relevant quantity: since $|W|$ is constant inside each $(n_h,n_l)$ sector, the perfect model must be approximately uniform over same-sign configurations within that sector. Any residual spread in $q_{\pm}(x\mid n_h,n_l)$, caused by accumulated next-token prediction errors, inflates $\mathrm{Var}(h)$ without improving its covariance with the sign. Figure~\ref{fig:logqlogw} confirms this picture: the strong linear relation between $\log q$ and $\log |W|$ shows that the coarse $(n_h,n_l)$ dependence is easily captured, while the finite width of the $\log(q/|W|)$ distribution reflects the within-sector modeling error that ultimately limits the variance reduction.

It is worth noting the gap between the training objective and the CV goal. The model is trained by minimizing the KL divergence $D_{\mathrm{KL}}(p_\sigma\|q_\sigma)$, which operates in log space and drives $\log q$ toward $\log |W|$ within each sign sector. The control variate $h$ [Eq.~(\ref{eq:hx})], however, depends on the ratio $q/|W|=\exp[\log(q/|W|)]$ in linear space. Because of this exponential map, a small residual error $\epsilon$ in $\log(q/|W|)$ translates into a multiplicative factor $e^{\epsilon}$ in $q/|W|$, which can be substantially amplified for configurations with large $|\epsilon|$. The $\log q$-vs-$\log |W|$ scatter plot (Fig.~\ref{fig:logqlogw}) is therefore a necessary but not sufficient diagnostic: it confirms that the log-space training objective is well optimized, but the CV variance reduction is sensitive to the tails of the $\log(q/|W|)$ distribution, where rare large deviations are exponentially magnified.

\section{Discussion and Outlook}
\label{sec:discussion}

This work develops a methodological framework for constructing neural control variates in quantum Monte Carlo simulations with sign problems. The key innovation is the use of autoregressive models with exact normalization and strict sign-sector resolution to provide structurally zero-mean control variates that reduce estimator variance while preserving unbiasedness. The autoregressive factorization guarantees that the control variate has zero mean by construction, eliminating the need for post-hoc bias correction. The benchmarks on small frustrated clusters serve as proof-of-principle demonstrations that the framework is conceptually sound and numerically effective in regimes where the sign problem is already severe. We do not claim that the current implementation is ready for production use on large systems. Rather, we establish the feasibility of the approach and identify the key architectural and algorithmic components required for future scaling efforts.

Benchmarks on the triangular-lattice Heisenberg antiferromagnet demonstrate the effectiveness of the method in the small-$N$ limit. On the 3-site lattice, the control variate achieves ${\sim}8$--$11\times$ SE improvement on $\avg{s}$ and ${\sim}2.2$--$5.3\times$ on energy across $\beta=6$--$15$. On the $2\times 2$ lattice (tetrahedron), ${\sim}5$--$9\times$ SE improvement on $\avg{s}$ and ${\sim}2.4$--$5\times$ on energy persist even as $\avg{s}$ drops below $10^{-3}$. On larger systems such as the $3\times 3$ lattice, the improvement is substantially reduced, with degradation reflecting fixed model capacity rather than a fundamental limitation.

Two bottlenecks limit the current variance reduction. The first is the gap between the training objective and the CV performance. The model is trained by minimizing $D_{\mathrm{KL}}(p_\sigma\|q_\sigma)$, which operates in log space, but the control variate involves the ratio $q/|W|$ in linear space. As discussed in Sec.~\ref{sec:results}, the exponential map amplifies small residual errors in $\log(q/|W|)$, so the CV is sensitive to the tails of the modeling-error distribution even when the log-space loss is well converged. The second bottleneck is the lack of structural inductive bias. In the current architecture, each bond index is treated as a plain token: all bond connectivity, vertex-list linkage, and loop topology must be inferred by the transformer from data alone. The MLP head ablation (Sec.~\ref{sec:results}) demonstrates this directly---explicitly injecting the candidate $\Delta n_l$ at the output yields a ${\sim}30$--$50\%$ relative gain in SE reduction, indicating that the transformer alone struggles to recover even this single graph-topological feature. Data augmentation techniques---cyclic shifts of the operator string and spatial symmetry operations---further improve the model by exposing it to equivalent representations of the same configuration, though their effect is less dramatic than the MLP head.

At a conceptual level, the control-variate framework transforms the sign problem---an exponential signal-to-noise difficulty intrinsic to the Monte Carlo estimator---into a neural-network learning problem. The variance reduction factor $1-\rho^2$ is controlled entirely by how well the learned distribution approximates the sign-resolved target, i.e., by the residual KL divergence $D_{\mathrm{KL}}(p_\sigma\|q_\sigma)$. Unlike the sign problem itself, which is a fixed property of the physical system and the chosen representation, the modeling error is a quantity that can be systematically reduced by improving the network architecture, increasing model capacity, or supplying better training data. The sign problem does not disappear, but it is recast into a form that admits a continuous resource--performance tradeoff: every improvement in the model translates directly into a larger $\rho$ and a smaller estimator variance.

The present benchmarks are limited to small systems where exact diagonalization is available. The Pearson correlation $\rho$ between the sign estimator and the control variate is remarkably stable across $\beta$, suggesting that the model quality does not degrade with the severity of the sign problem per se, but the $3\times 3$ results show reduced improvement factors, reflecting the increased difficulty of modeling both longer operator strings and a larger vocabulary in the expanded configuration space. This degradation is expected in the current fixed-capacity setup: the model capacity ($d_{\text{model}}=128$, $L=4$) is held constant across all system sizes, so the per-configuration modeling budget shrinks exponentially as the configuration space grows. Scaling the model capacity in tandem with the system size---a standard requirement in neural-network applications---is necessary to maintain a constant modeling quality per configuration.

Beyond scaling model capacity, physics-informed architectures offer a more fundamental direction for improvement. The current transformer treats the operator string as a flat token sequence without explicit knowledge of lattice geometry, which means the model must learn all spatial structure from data. Architectures that encode the graph structure of the QMC simulation cell directly---such as graph neural networks or message-passing layers that operate on the vertex graph---could supply the bond connectivity and loop topology as built-in features rather than quantities to be learned. This would reduce the effective complexity of the learning target and potentially steepen the scaling of model quality with network size, in analogy with the neural scaling laws observed in language modeling~\cite{Kaplan2020}. We view the identification of architectures whose inductive bias matches the structure of the target physical system---rather than brute-force enlargement of generic models---as the central challenge for scaling the method to larger lattices. Establishing the system-size scaling of the variance reduction under capacity-matched conditions---and whether it can keep pace with the exponential growth of the sign problem---is the critical open question. Our future work will focus on developing physics-informed architectures, such as graph neural networks that encode the lattice connectivity and loop topology directly, to address this scaling challenge.

The construction presented in this paper generalizes beyond the SU(2) Heisenberg model. For an anisotropic spin-$1/2$ XXZ model, diagonal and off-diagonal operator weights differ, so the equal-weight \RSSE\ resummation used above does not apply directly. Nevertheless, the incremental update scheme can be adapted: operators are inserted or removed one at a time, with the topological channel providing the corresponding changes in operator counts. The equal-weight loop coloring is then replaced by a generalized non-equal-weight loop assignment, producing a colored configuration that maps one-to-one onto an uncolored operator string plus a loop-spin assignment. The autoregressive model acts on the compact input $(\text{uncolored }S_M,\,\boldsymbol{\sigma}_{\mathrm{loop}})$ and can learn the resulting weight asymmetry through its conditional distributions.

More broadly, the underlying principle---constructing a zero-mean control variate from an exactly normalized generative model---applies to any Monte Carlo framework in which configurations can be represented as discrete sequences. The essential requirement is \emph{exact normalization} over the physical configuration space, guaranteed by a masking scheme that assigns zero probability to nonphysical configurations. This ensures that the control variate has zero expectation under the sampling measure, thereby preserving unbiasedness.

In the present SSE implementation, we further impose \emph{exact sign-sector resolution}: each model's support is strictly confined to one parity sector by the end-of-sequence mask. This separation brings a practical advantage---the model only needs to learn the amplitude structure within its sector, rather than simultaneously learning both amplitude and sign structure, resulting in improved generalization and training stability. However, this separation is not a fundamental requirement. In principle, one could train $q_+(x)$ and $q_-(x)$ without strict sector confinement: each configuration would be evaluated by both models, and cross-entropy training would drive each model to assign high probability to samples from its training sector and low probability to the opposite sector (provided the training data are still partitioned by sign). The normalization and zero-expectation properties would still hold, but the models would need to learn a more complex joint distribution over amplitude and sign, potentially at the cost of generalization. For determinant QMC, where the sign is readily computable (via a determinant evaluation) but difficult to impose structurally through masking, this relaxed approach may be more practical. Beyond the question of sign-sector separation, the challenge in applying this framework to other QMC methods lies in designing model architectures whose inductive bias matches the structure of the target physical system and its configuration-space representation.

\noindent\textbf{Code availability.}
Code and data are available at \url{https://github.com/Joe-Nor/NCV_for_QMC}.

\begin{acknowledgments}
The authors acknowledge valuable discussions with Yizhuang You and Yuan Wan. This work is supported by the National Natural Science Foundation of China under Grants No. T2225018, No. 12188101, No. T2121001, the Cross-Disciplinary Key Project of Beijing Natural Science Foundation No. Z250005, the Strategic Priority Research Program of the Chinese Academy of Sciences under Grants No. XDB0500000, and the National Key Projects for Research and Development of China Grants No. 2021YFA1400400.
\end{acknowledgments}

\appendix

\section{Zero-Variance Analysis for the Control Variates}
\label{app:zerovar}

\subsection{Denominator: exact zero variance}

In the perfect-model limit, the control variate $h(x)$ defined in Eq.~(\ref{eq:hx}) takes the sectorwise-constant form
\begin{equation}
  h(x)=
  \begin{cases}
    1/Z_+, & s(x)=+1, \\
    -1/Z_-, & s(x)=-1.
  \end{cases}
\end{equation}
Defining $p_{\pm}=Z_{\pm}/Z$, one has $\avg{s}=p_+-p_-$ and $\mathrm{Var}(s)=4p_+p_-$. The second moment of $h$ is
\begin{equation}
  \avg{h^2}
  =p_+\frac{1}{Z_+^2}+p_-\frac{1}{Z_-^2}
  =\frac{1}{ZZ_+}+\frac{1}{ZZ_-}
  =\frac{1}{Z_+Z_-},
\end{equation}
so $\mathrm{Var}(h)=1/(Z_+Z_-)$ (since $\avg{h}=0$ by Eq.~(\ref{eq:ehzero})). The cross moment is
\begin{equation}
  \avg{s\,h}
  =p_+\frac{1}{Z_+}+p_-\frac{1}{Z_-}
  =\frac{1}{Z}+\frac{1}{Z}
  =\frac{2}{Z},
\end{equation}
giving $\mathrm{Cov}(s,h)=2/Z$. The optimal coefficient is therefore
\begin{equation}
  c^*=\frac{\mathrm{Cov}(s,h)}{\mathrm{Var}(h)}
  =\frac{2/Z}{1/(Z_+Z_-)}
  =\frac{2Z_+Z_-}{Z}.
\end{equation}
Substituting into the variance formula,
\begin{align}
  \mathrm{Var}(s-c^*h)
  &=\mathrm{Var}(s)-\frac{\mathrm{Cov}(s,h)^2}{\mathrm{Var}(h)} \nonumber\\
  &=4p_+p_--\frac{4/Z^2}{1/(Z_+Z_-)}
  =4\frac{Z_+Z_-}{Z^2}-4\frac{Z_+Z_-}{Z^2}=0.
\end{align}
One can verify this configuration by configuration. For $s(x)=+1$,
\begin{equation}
  s-c^*h
  =1-\frac{2Z_+Z_-}{Z}\frac{1}{Z_+}
  =1-\frac{2Z_-}{Z}
  =p_+-p_-,
\end{equation}
and for $s(x)=-1$,
\begin{equation}
  s-c^*h
  =-1+\frac{2Z_+Z_-}{Z}\frac{1}{Z_-}
  =-1+\frac{2Z_+}{Z}
  =p_+-p_-.
\end{equation}
Hence each individual sample already returns the exact average sign $\avg{s}$.

\subsection{Numerator: approximate cancellation}

For the numerator $\avg{n_h\,s}$, the $n_h$-weighted models yield in the perfect limit
\begin{equation}
  h_O(x)=
  \begin{cases}
    n_h(x)/Z_+^{(O)}, & s(x)=+1, \\[2pt]
    -n_h(x)/Z_-^{(O)}, & s(x)=-1,
  \end{cases}
\end{equation}
where $Z_\pm^{(O)}=\sum_{x:s=\pm 1}n_h(x)|W(x)|$. The target observable is $y(x)=n_h(x)\,s(x)$. For exact zero variance one would need a constant $\lambda$ such that
\begin{equation}
  y(x)-\lambda\,h_O(x)=\mathrm{const}
\end{equation}
for every configuration $x$. Inside the positive sector this becomes
\begin{equation}
  n_h(x)\bigl(1-\lambda/Z_+^{(O)}\bigr)=\mathrm{const},
\end{equation}
and inside the negative sector
\begin{equation}
  -n_h(x)\bigl(1-\lambda/Z_-^{(O)}\bigr)=\mathrm{const}.
\end{equation}
Since $n_h(x)$ fluctuates within each sign sector, both prefactors must vanish, which requires
\begin{equation}
  \lambda=Z_+^{(O)}=Z_-^{(O)}.
\end{equation}
This equality is non-generic, so the numerator CV has no exact zero-variance limit.

Nevertheless, the residual variance can be small. Writing $Z_\pm^{(O)}=\bar{Z}^{(O)}(1\pm\delta)$ with a small asymmetry parameter $\delta$ and choosing $\lambda=\bar{Z}^{(O)}$,
\begin{equation}
  y-\lambda\,h_O=
  \begin{cases}
    n_h\,\delta + O(\delta^2), & s=+1, \\
    n_h\,\delta + O(\delta^2), & s=-1,
  \end{cases}
\end{equation}
up to the sector-dependent expansion of $1/(1\pm\delta)$. The leading cancellation survives, and the remaining variance is controlled by $\delta$---the mismatch between the $n_h$-weighted moments of the two sign sectors. When the two sectors have similar $n_h$ distributions, only higher-order residual terms remain.

\section{Sign-sector ergodicity: autocorrelation functions}
\label{app:acf}

\begin{figure}[t]
\centering
\includegraphics[width=\columnwidth]{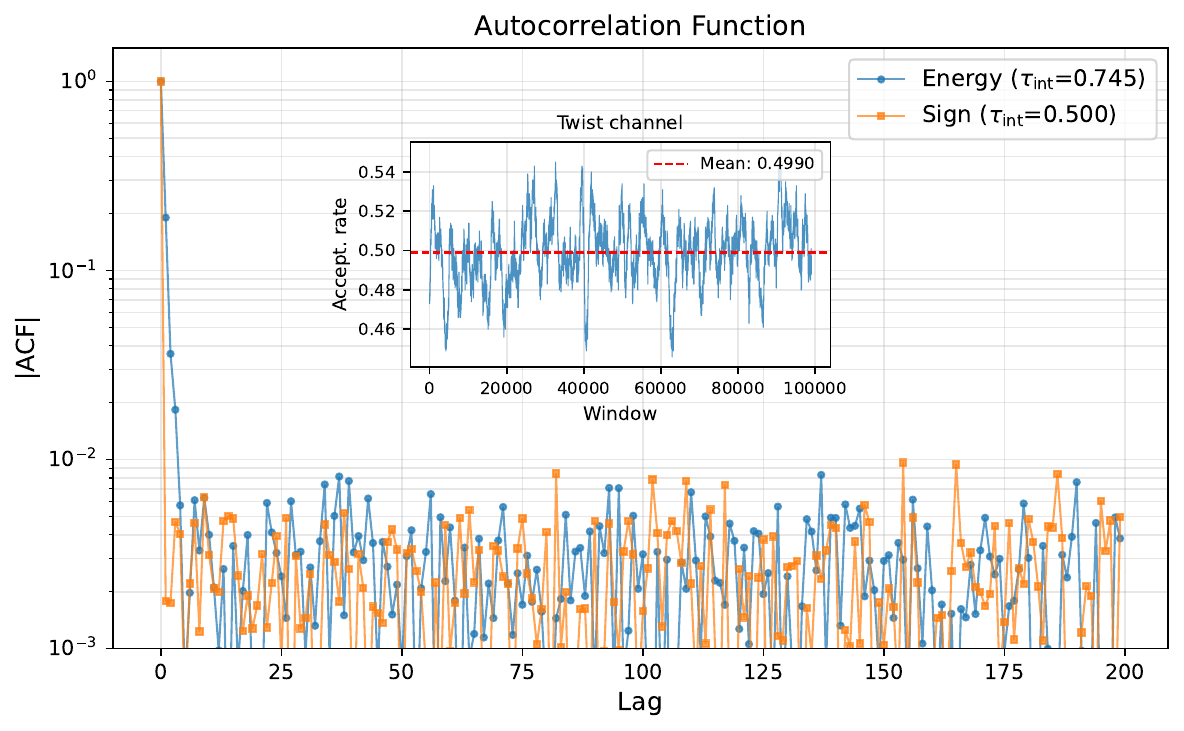}
\caption{\label{fig:acf}
Autocorrelation functions of the energy (blue) and the sign (orange) on the $3\times 3$ triangular lattice at $\beta=3$. The sign ACF decorrelates within a single sweep, rapidly reaching the $10^{-3}$ level, confirming rapid sign-sector mixing through the twist channel. The energy ACF decorrelates within ${\sim}10$ sweeps. One sweep consists of sequentially visiting all $M$ positions and proposing insertions or removals.}
\end{figure}

To verify that the \RSSE\ Markov chain with the incremental loop-topology update provides ergodic sampling across both sign sectors, we measure the autocorrelation functions of the sign and the energy. Figure~\ref{fig:acf} shows the results for the $3\times 3$ triangular lattice at $\beta=3$, where the average sign is $\avg{s}\approx 1.0\times 10^{-4}$ and the sign problem is extremely severe.

The autocorrelation function is defined as
\begin{equation}
  C_A(\tau) = \frac{\langle A(t)\,A(t+\tau)\rangle - \langle A\rangle^2}{\langle A^2\rangle - \langle A\rangle^2},
\end{equation}
where $A$ is either the sign $s(x)$ or the energy $E(x)$, and $\tau$ is measured in units of MCMC sweeps. One sweep consists of sequentially visiting all $M$ positions in the operator string and proposing either an insertion or a removal at each position.

The sign autocorrelation function decorrelates within a single sweep, rapidly reaching the $10^{-3}$ level, confirming that the twist channel provides rapid mixing between sign sectors. The energy autocorrelation decorrelates within ${\sim}10$ sweeps, indicating that the overall Markov chain equilibrates efficiently. The rapid decorrelation of the sign demonstrates that the chain is ergodic and that both sign sectors are visited with the correct equilibrium probabilities.

The twist channel statistics provide further insight into the sign-sector mixing mechanism. Over $8.2\times 10^6$ total update attempts, twist-channel moves (those with $\Delta n_l=0$) account for approximately $10\%$ of all proposals and are accepted at a rate of ${\sim}75\%$, substantially higher than the overall acceptance rate of ${\sim}59\%$. This high twist acceptance rate, combined with the rapid sign decorrelation, confirms that the incremental loop-topology update provides efficient ergodic sampling across both sign sectors even in frustrated systems where the sign problem is severe.

\FloatBarrier
\bibliography{references}

\end{document}